\documentclass[journal,preprint]{vgtc}

\PassOptionsToPackage{x11names,svgnames}{xcolor}

\newif\ifEngVer
\EngVertrue

\usepackage{environ}

\newtoggle{engver}
\ifEngVer \toggletrue{engver} \else \togglefalse{engver} \fi

\ifEngVer
\newenvironment{EN}{}{}
\NewEnviron{JA}{}
\else
\newenvironment{JA}{}{}
\NewEnviron{EN}{}
\fi
\onlineid{0}

\vgtccategory{Research}

\title{DELUGE: Decomposed Entropy-coded Live Unstructured Geometry Exchange for Real-time Particle Streaming}

\author{  \authororcid{Hikari Yanagawa}{0009-0003-4539-6432},
  \authororcid{Yuichi Hiroi}{0000-0001-8567-6947}, and
  \authororcid{Takefumi Hiraki}{0000-0002-5767-3607}
}

\authorfooter{
  \item
  Hikari Yanagawa is with Cluster Metaverse Lab and University of Tsukuba.
  E-mail: h.yanagawa@cluster.mu.
  \item
  Yuichi Hiroi is with Cluster Metaverse Lab.
  E-mail: y.hiroi@cluster.mu.
  \item
  Takefumi Hiraki is with University of Tsukuba and Cluster Metaverse Lab.
  E-mail: hiraki@slis.tsukuba.ac.jp.
}

\abstract{  Particle-based physics simulations, including fluids, smoke, and granular media, are fundamental to visual realism in immersive VR and AR.
  With the growing adoption of social VR and digital twins, demand is increasing for shared experiences in which multiple users interact with the same simulation in real time.
  Realizing such experiences requires low-latency streaming of large-scale particle data from a server to each client, yet existing point cloud compression methods such as G-PCC (TMC13) and Draco assume static geometric structures; when applied to dynamic particle streaming, their encoding latency exceeds the frame period, failing to meet real-time delivery requirements.
  We propose DELUGE, a streaming compression architecture that exploits the temporal coherence and velocity predictability inherent in physics simulation particles, achieving sub-frame-latency encoding and decoding through three complementary techniques.
  Evaluation on dynamic point cloud datasets demonstrates that DELUGE achieves approximately $20\times$ faster decoding than G-PCC (TMC13) and $6\times$ faster encoding than Draco.
  We further build end-to-end client implementations for both web browsers and Apple Vision Pro, and confirm through a within-participants perceptual quality evaluation and a two-person collaborative task study on Vision Pro that the proposed method supports real-time collaborative experiences with hand-tracked fluid interaction.
}


\teaser{
  \centering
  \includegraphics[width=\linewidth]{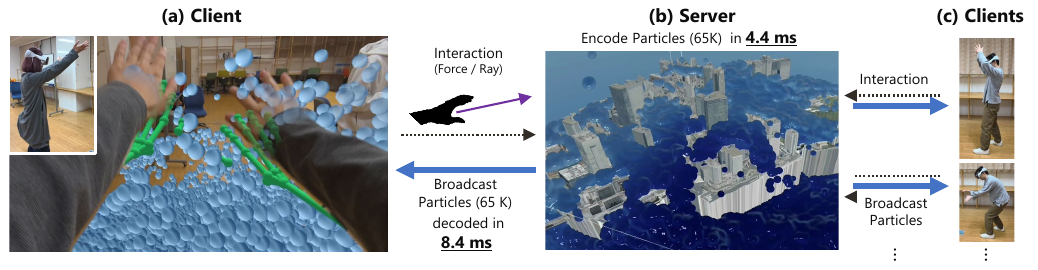}
  \caption{DELUGE enables real-time multi-user interaction with streaming particle simulations. (a)~A user on Apple Vision Pro manipulates 65K fluid particles with bare-hand tracking. (b)~The server runs an MLS-MPM simulation and encodes each keyframe in 4.4\,ms via the DELUGE codec. The compressed stream is broadcast to all connected clients via WebSocket. (c)~The encoded packet is broadcast to an arbitrary number of clients without additional encoding cost; each client independently decodes it in 8.4\,ms on the Vision Pro device and interacts simultaneously, with hand inputs reflected back into the shared simulation.}
  \label{fig:teaser}
  \vspace{-1mm}
}

\graphicspath{{figures/}{pictures/}{images/}{./}}

\usepackage{booktabs}
\usepackage{tabularray}
\UseTblrLibrary{booktabs}
\usepackage{mathptmx}
\usepackage{amsmath,multirow,siunitx,amssymb}
\usepackage{tabularx}
\usepackage[whole]{bxcjkjatype}
\usepackage{balance}
\usepackage[mathcal]{eucal}
\usepackage[most]{tcolorbox}
\usepackage{multicol}
\usepackage{xspace}
\usepackage{soul}
\usepackage{docmute}

\newcommand{\etal}{et al.\@\xspace}

\newcommand{\eg}{e.g.\@\xspace}

\newcommand{\ie}{i.e.\@\xspace}

\newcommand{\commentout}[1]{}

\begin{document}

  \firstsection{Introduction}
  \maketitle
  \label{sec:introduction}

  \begin{JA}
    粒子ベース物理シミュレーション（流体・煙・砂など）は，没入型VR・AR体験の視覚的リアリズム向上~\cite{muller_particle-based_2003}，デジタルツインを活用した都市洪水可視化~\cite{tsujimoto2024floodmr,ford_digital_twin_flood_2020}，産業用流体解析~\cite{diewald_challenges_2022}など多様な領域で不可欠な技術である．近年のソーシャルVR環境の普及~\cite{han_people_places_2023}により，複数ユーザが同一の物理現象に双方向にインタラクションする共有体験への需要が高まっている~\cite{mcveighschultz_prosocial_2019}．こうした共有物理体験は共同存在感（co-presence）の向上~\cite{kyrlitsias_social_presence_2022}のみならず，地理的に分散した関係者間のリアルタイム協調を可能にする~\cite{ibara_data_integration_2025}．しかしその実現には，大規模粒子データを低遅延で複数クライアントに配信しつつ，各クライアントからの入力を即座にシミュレーションに反映する必要がある．
  \end{JA}
  \begin{EN}
    Particle-based physics simulations, including fluids, smoke, and granular media, are indispensable across a range of domains: improving visual realism in immersive VR and AR~\cite{muller_particle-based_2003}, visualizing urban flooding through digital twins~\cite{tsujimoto2024floodmr,ford_digital_twin_flood_2020}, and industrial fluid analysis~\cite{diewald_challenges_2022}, among other applications. The growing adoption of social VR platforms~\cite{han_people_places_2023} has increased the demand for shared experiences in which multiple users can interact bidirectionally with the same physical phenomena~\cite{mcveighschultz_prosocial_2019}. These shared physics experiences improve co-presence~\cite{kyrlitsias_social_presence_2022} and enable real-time collaboration among geographically distributed participants~\cite{ibara_data_integration_2025}. However, realizing these experiences requires delivering large-scale particle data to multiple clients at low latency while reflecting each client's input into the simulation without delay.
  \end{EN}

  \begin{JA}
    この要件に対し，既存のアプローチはいずれも本質的な制約を抱えている．クライアント側でシミュレーションを実行する手法~\cite{muller_particle-based_2003}はモバイル端末の計算能力に強く制約され，大規模粒子系の処理が困難である．クラウドゲーミングに代表されるビデオストリーミング方式はクライアント負荷を低減できるが，ユーザの身体運動から視覚更新までの遅延（Motion-to-Photon遅延，以下M2P遅延）がインタラクションの即応性を損なう~\cite{gul_low-latency_2020}．一方，ジオメトリストリーミングでは，クライアントが最新のジオメトリを保持し，頭部運動による視点変化をサーバ更新を待たずにローカルで再描画できる．ただし，シミュレーション状態の更新には引き続きサーバとの通信が必要である．しかし，MPEG G-PCC~\cite{schwarz_emerging_2019}等のOctreeベースのintra符号化は静的な点群構造を前提とした空間分割に基づいており，    パーティクル個々の速度を利用しない．そのため動的粒子のストリーミングではフレームごとに空間構造を再構築する必要が生じ，符号化レイテンシがフレーム周期を超過する．
  \end{JA}
  \begin{EN}
    Existing approaches all face fundamental limitations in meeting this requirement. Running the simulation on the client~\cite{muller_particle-based_2003} is heavily constrained by the computational capacity of mobile devices, making large-scale particle processing infeasible.
    Although video streaming approaches such as cloud gaming reduce client workload, they introduce motion-to-photon (M2P) latency, the delay from a user's physical motion to the corresponding visual update, that degrades interaction responsiveness~\cite{gul_low-latency_2020}. By contrast, geometry streaming lets the client retain the latest geometry and re-render head-motion-induced viewpoint changes locally without waiting for the server, while simulation-state updates still require server communication. This makes the point cloud codec the latency-critical component, however, octree-based intra coders such as MPEG G-PCC~\cite{schwarz_emerging_2019} rely on spatial partitioning designed for static point distributions and do not exploit the per-particle velocity inherent in physics simulation particles.     Consequently, applying these standards to dynamic particle streaming requires reconstructing the spatial structure at every frame, resulting in encoding latency that exceeds the frame period.
  \end{EN}

  \begin{JA}
    本研究では，動的粒子データが持つ時間コヒーレンスと速度予測可能性——既存コーデックが低遅延インタラクティブ復号のために十分に活用していない性質——に着目し，ストリーミング圧縮アーキテクチャ\textbf{DELUGE} (Decomposed Entropy-coded Live Unstructured Geometry Exchange)を提案する．DELUGEは，MPEG映像圧縮~\cite{wiegand_overview_2003,sullivan_overview_2012}におけるキーフレーム/デルタフレーム構造を3次元粒子データに拡張し，3つの相補的な手法を組み合わせる．(i) \textbf{速度適応型ビット配分}は各パーティクルの速度に応じてOctree深度を動的に制御し，量子化残差分布をリーフ間で均一化する．(ii) \textbf{軸分離型エントロピー符号化}は量子化された$x$, $y$, $z$のシンボル列を独立したストリームとして符号化し，軸固有の統計構造を活用しつつ3軸の完全並列復号を可能にする．(iii) \textbf{平坦逆量子化LUT}はリーフ単位の逆量子化パラメータをパーティクル単位の平坦配列に事前展開し，木の走査を$O(N)$の単一ループ再構成に置き換える．これら3手法の統合により，符号化・復号の両レイテンシをフレーム周期未満に抑える．  \end{JA}
  \begin{EN}
    We focus on temporal coherence and velocity predictability, two properties inherent in dynamic particle data that prior codecs leave underused for low-latency interactive decoding. We propose \textbf{DELUGE} (Decomposed Entropy-coded Live Unstructured Geometry Exchange), a streaming compression architecture that extends the keyframe/delta-frame structure of MPEG video compression~\cite{wiegand_overview_2003,sullivan_overview_2012} to three-dimensional particle data. DELUGE combines three complementary techniques: (i) \textbf{Velocity-adaptive bit allocation} dynamically controls the Octree depth according to the velocity of each particle, which equalizes the quantization residual distribution across leaves. (ii) \textbf{Axis-separated entropy coding} encodes the quantized $x$, $y$, and $z$ symbol sequences as independent streams, exploiting axis-specific statistical structure while enabling fully parallel three-axis decoding. (iii) \textbf{A flat inverse-quantization look-up table (LUT)} pre-expands per-leaf dequantization parameters into a per-particle flat array, replacing tree traversal with $O(N)$ single-loop reconstruction. Together, these techniques reduce both encoding and decoding latency below the frame period.  \end{EN}

  \begin{JA}
    DELUGEはRD効率ではなくデコードレイテンシを最適化対象として設計されている．動的点群データセットを用いた評価において，DELUGEはG-PCC（TMC13）に対し復号速度を約20倍改善し，Dracoに対しては符号化速度を6倍改善した（全コーデックFFI in-process計測）．

    さらに，WebブラウザおよびApple Vision Proの両プラットフォームにおけるクライアント実装を構築し，Vision Pro上でのハンドトラッキングによる流体インタラクションを用いた被験者内知覚品質評価と2者間協調タスク評価により，提案手法がリアルタイムの協調体験を実現することを確認した．
  \end{JA}
  \begin{EN}
    DELUGE is designed to optimize decoding latency rather than rate-distortion efficiency. In evaluations on dynamic point cloud datasets, DELUGE achieves approximately $20\times$ faster decoding than G-PCC (TMC13) and $6\times$ faster encoding than Draco, when all codecs are measured in-process via FFI.
    We further build end-to-end client implementations for web browsers and Apple Vision Pro, and confirm through a within-participants perceptual quality evaluation and a two-person collaborative task study on Vision Pro that the proposed method enables real-time collaborative experiences with hand-tracked fluid interaction.
  \end{EN}

  \begin{JA}
    本研究の主な貢献は以下の通りである．
  \end{JA}
  \begin{EN}
    The main contributions of this work are as follows:
  \end{EN}
  \begin{JA}
    \begin{itemize}[leftmargin=*]
  \item 物理シミュレーション粒子の時間コヒーレンスと速度予測可能性をコーデックパイプラインの全段階で活用するストリーミング圧縮手法DELUGE．
  \item 速度適応型ビット配分，軸分離型エントロピー符号化，平坦逆量子化LUTの3技術の設計と統合により，符号化効率と復号スループットを同時に改善．
  \item WebブラウザおよびApple Vision Pro（ハンドトラッキング対応）を含むマルチプラットフォームのエンドツーエンドストリーミング実装．
  \item G-PCC/Dracoとのレート歪み比較，主観評価による知覚品質評価，没入環境における2者間協調タスク評価を組み合わせた包括的な評価．
    \end{itemize}
  \end{JA}
  \begin{EN}
    \begin{itemize}[leftmargin=*]
  \item DELUGE, a streaming compression method that exploits the temporal coherence and velocity predictability of physics simulation particles across every stage of the codec pipeline.
  \item The design and integration of three techniques, velocity-adaptive bit allocation, axis-separated entropy coding, and a flat inverse-quantization LUT, that jointly improve coding efficiency and decoding throughput.
  \item Multi-platform end-to-end streaming client implementations, including a web browser client and an Apple Vision Pro client with hand tracking.
  \item A comprehensive evaluation combining rate-distortion comparison with G-PCC and Draco, perceptual quality assessment via subjective testing, and a two-user collaborative task study in an immersive environment.
    \end{itemize}
  \end{EN}

  \section{Related Work}\label{sec:related}

  \subsection{Point Cloud and Particle Interaction in VR/AR}

  \begin{JA}
    粒子ベース流体シミュレーションの研究は，Müllerら~\cite{muller_particle-based_2003}によるSPHを端緒として発展してきた．GPU並列化の進展により数十万粒子規模のリアルタイム処理が実現され，近年ではHuら~\cite{hu_moving_2018}のMLS-MPMが近傍探索を不要とする特性から注目を集め，Webブラウザ上でも大規模シミュレーションを可能としている．しかし，シミュレーション結果を複数クライアントへリアルタイム配信する技術は未だ確立されていない．
  \end{JA}
  \begin{EN}
    Particle-based fluid simulation has evolved considerably since M\"{u}ller~\etal~\cite{muller_particle-based_2003} introduced interactive SPH (Smoothed Particle Hydrodynamics). GPU parallelism now enables real-time processing of hundreds of thousands of particles, and MLS-MPM by Hu~\etal~\cite{hu_moving_2018} has broadened the design space further by eliminating neighbor searches, making large-scale simulations feasible even within web browsers. Nevertheless, techniques for delivering such simulation results to multiple remote clients in real time remain largely unexplored.
  \end{EN}

  \begin{JA}
    VR/AR環境における流体可視化~\cite{zhibin2022airflow,cen2024realtime-fluid}やデジタルツインを活用した洪水リスク評価~\cite{ford_digital_twin_flood_2020,tsujimoto2024floodmr}も進展しているが，いずれも単一ユーザまたは事前計算結果の可視化が主であり，リアルタイムシミュレーションへの双方向介入には至っていない．動的粒子群を毎秒60フレームで複数ユーザ間にリアルタイム配信・操作する枠組みは未だ存在しない．
  \end{JA}
  \begin{EN}
    Fluid visualization in VR/AR~\cite{zhibin2022airflow,cen2024realtime-fluid} and digital-twin-based flood risk assessment~\cite{ford_digital_twin_flood_2020,tsujimoto2024floodmr} have also progressed, but these efforts target single-user settings or precomputed results and do not support bidirectional user intervention in a live simulation. No existing framework enables multiple users to simultaneously observe and manipulate dynamic particle fields streamed at interactive rates.
  \end{EN}

  \subsection{Point Cloud and Volumetric Content Compression}

  \begin{JA}
    点群圧縮の標準化において，MPEGはV-PCC~\cite{vpcc2020}とG-PCC~\cite{gpcc2020}の2方式を策定している~\cite{schwarz_emerging_2019}．V-PCCは点群を2D映像に射影して圧縮するため，復号後の個別点アクセスが困難でありインタラクティブ用途に適さない．G-PCCはOctreeベースの空間符号化~\cite{meagher_geometric_1982}で3次元構造を直接符号化するが，動的粒子のストリーミングではフレームごとにOctreeを再構築する必要があり符号化レイテンシが増大する．Google Draco~\cite{draco_github}等の軽量コーデックも単一フレームの静的点群を対象としている．動的点群では，V-PCCが2D映像コーデックを介して動き情報を利用する一方，de QueirozとChou~\cite{dequeiroz2017motion}はボクセル化点群のブロック単位動き補償圧縮（MCIC）を提案している．MCICは動き推定を対象外として外部RGBDパイプラインの対応付けに依存し，アーティファクト修復を要する非可逆幾何符号化のMATLABプロトタイプで符号化時間も報告していない．これに対し，DELUGEはシミュレータが提供するIDと連続座標を利用して動き推定と修復処理を不要とし，Webの60\,fpsおよびVision Proの90\,fpsの時間制約を満たし，高ビットレートではほぼロスレスの品質に到達する．  \end{JA}
  \begin{EN}
    MPEG has standardized two point cloud compression approaches~\cite{schwarz_emerging_2019}: V-PCC~\cite{vpcc2020}, which projects points onto 2D video and thus precludes per-point random access needed for interactive manipulation, and G-PCC~\cite{gpcc2020}, which encodes 3D geometry via octree-based spatial partitioning~\cite{meagher_geometric_1982} but requires reconstructing the octree at every frame when applied to dynamic particles, incurring substantial encoding latency. Lightweight single-frame codecs such as Google Draco~\cite{draco_github} also target static point clouds. For dynamic point clouds, V-PCC exploits motion through 2D video codecs, while de Queiroz and Chou~\cite{dequeiroz2017motion} proposed block-wise motion-compensated compression of voxelized point clouds (MCIC). MCIC excludes motion estimation, relying on an external RGBD pipeline for correspondences, and evaluates a lossy MATLAB prototype requiring artifact repair without reporting encoding time. DELUGE instead uses simulator-provided IDs and continuous coordinates, avoiding motion estimation and repair; it meets the 60\,fps Web and 90\,fps Vision Pro budgets and reaches near-lossless quality at high rates.  \end{EN}

  \begin{JA}
    深層学習を活用した点群圧縮も活発に研究されており，VoxelDNN~\cite{nguyen_learning-based_2020}はG-PCCを上回る圧縮効率を達成した．動的点群に対してはニューラルモーション補償~\cite{guarda_motion_2020}，D-DPCC~\cite{fan_ddpcc_2022}，HINT~\cite{gao_hint_2026}が提案されている．しかし，これらの手法は推論にGPU上で数百ms〜数秒を要し（例：D-DPCCは1フレームあたり約1.7秒~\cite{fan_ddpcc_2022}），60\,fpsのフレーム周期16.67\,msとは桁が異なるため，リアルタイム復号の対象とならない．本研究では軽量な復号を維持しつつ時間的コヒーレンスを活用するため，伝統的なエントロピー符号化と動的粒子データ固有の統計的特性を組み合わせるアプローチを採用する．
  \end{JA}
  \begin{EN}
    Learning-based point cloud compression has also received considerable attention. VoxelDNN~\cite{nguyen_learning-based_2020} surpasses G-PCC in compression efficiency. For dynamic point clouds, neural motion compensation~\cite{guarda_motion_2020}, D-DPCC with end-to-end motion estimation and compensation~\cite{fan_ddpcc_2022}, and hierarchical inter-frame correlation~\cite{gao_hint_2026} have been proposed. However, these methods require hundreds of milliseconds to seconds of GPU inference per frame (\eg D-DPCC reports $\sim$1.7\,s/frame~\cite{fan_ddpcc_2022}), which is orders of magnitude above the 16.67\,ms budget at 60\,fps, precluding real-time decoding. To achieve lightweight decoding while exploiting temporal coherence, we combine traditional entropy coding with statistical properties specific to dynamic particle data instead of relying on learned models.
  \end{EN}

  \begin{JA}
    6DoF体験のための点群ストリーミングでは，HTTP適応ストリーミング（DASH）の統合~\cite{van_der_hooft_towards_2019}，V-PCCストリームのライブトランスコーディング（RABBIT）~\cite{rudolph_rabbit_2023}，パッチベースの3D圧縮フレームワーク~\cite{liu_patchvvc_2023}が提案されている．しかし，これらは受動的視聴体験の品質適応に焦点を当てており，本研究が目指すユーザからの能動的介入を即座に反映する双方向ストリーミングとは目的が異なる．
  \end{JA}
  \begin{EN}
    For 6 degrees-of-freedom (DoF) point cloud streaming experiences, integration with HTTP adaptive streaming (DASH)~\cite{van_der_hooft_towards_2019}, live transcoding of V-PCC streams (RABBIT)~\cite{rudolph_rabbit_2023}, and patch-based 3D compression frameworks~\cite{liu_patchvvc_2023} have been proposed.
    While these efforts focus on quality adaptation for passive viewing, this work targets bidirectional streaming, wherein user interventions are immediately reflected in the shared simulation state.
  \end{EN}

  \subsection{Cloud-Based Graphics and Low-Latency Interaction}

  \begin{JA}
    リモートレンダリング~\cite{shi_survey_2015}はサーバ側でレンダリングした結果をクライアントに配信する方式であり，NVIDIA CloudXR~\cite{nvidia_cloudxr}等の商用サービスも提供されている．この手法はクライアントの計算負荷を最小化できる一方，ビデオストリームとしての配信が前提であり，視点変更時のM2P遅延の制約から逃れられない~\cite{cai_survey_cloud_gaming_2016}．
  \end{JA}
  \begin{EN}
    Remote rendering~\cite{shi_survey_2015} delivers server-side rendered output to clients, and commercial services such as NVIDIA CloudXR~\cite{nvidia_cloudxr} are available. While this approach minimizes client-side computation, it relies on video stream delivery and therefore cannot eliminate M2P latency upon viewpoint changes~\cite{cai_survey_cloud_gaming_2016}.
  \end{EN}

  \begin{JA}
    M2P遅延と画像品質を両立させるため，頭部運動予測による遅延補償~\cite{gul_low-latency_2020}やニューラルレンダリングのリモート実行とストリーミングの統合~\cite{hiroi_nearportation_2022}が提案されている．一方，LiveRender~\cite{liverender_mm_2014}はビデオストリーミングとは異なり，描画コマンドやジオメトリ情報を送信することで帯域削減と応答遅延の改善を実現した．本研究は「粒子状態を送信しクライアントでレンダリングする」方針においてLiveRenderと共通するが，動的粒子データの時間的コヒーレンスを速度適応型空間分割により活用する点で独自性がある．
  \end{JA}
  \begin{EN}
    To reconcile low M2P latency with high visual quality, head motion prediction for latency compensation~\cite{gul_low-latency_2020} and integration of remote neural rendering with streaming~\cite{hiroi_nearportation_2022} have been proposed. Taking a different approach, LiveRender~\cite{liverender_mm_2014} transmits drawing commands and geometry instead of video frames, thereby reducing bandwidth and response latency. While our work and LiveRender both transmit scene state for client-side rendering, our approach differs in that it exploits the temporal coherence of dynamic particle data through velocity-adaptive spatial partitioning.
  \end{EN}

  \begin{JA}
    Social VRプラットフォームにおいて複数ユーザ間のリアルタイム同期は体験品質の核心であり~\cite{han_people_places_2023,mcveighschultz_social_vr_2018}，共同存在感の形成~\cite{han_bailenson_social_vr_2024}や物理現象の共有による体験者の理解や認知の深化~\cite{ibara_data_integration_2025}が期待される．本システムは，複数ユーザが同一の流体シミュレーションへ同時に介入し即座に結果を共有する枠組みを提供することで，従来のSocial VRでは困難であった物理的協調作業を実現する．
  \end{JA}
  \begin{EN}
    On social VR platforms, real-time synchronization among users is central to the quality of the experience~\cite{han_people_places_2023,mcveighschultz_social_vr_2018}, which is fundamental to the formation of co-presence~\cite{han_bailenson_social_vr_2024} and supports deeper understanding through shared observation of physical phenomena~\cite{ibara_data_integration_2025}. Our system provides a framework in which multiple users can interact with the same fluid simulation and observe the results together in real time, enabling collaborative physical tasks that were previously impossible in conventional social VR settings.
  \end{EN}

  \section{Proposed Algorithm}\label{sec:proposed}

  \begin{JA}
    本章では，大規模パーティクルシミュレーションの低遅延ストリーミングを実現する圧縮アルゴリズムDELUGEを詳述する．DELUGEはビデオ圧縮~\cite{wiegand_overview_2003,sullivan_overview_2012}のキーフレーム/デルタフレーム構造を3次元粒子データに適用する．キーフレーム（ビデオ圧縮のI-frameに相当）は完全な空間構造と絶対量子化座標を送信し，デルタフレーム（P-frameに相当）は前フレームからの差分のみを送信する．
  \end{JA}
  \begin{EN}
    This section introduces the DELUGE compression algorithm, which enables the low-latency streaming of large-scale particle simulations. DELUGE uses the keyframe/delta-frame structure, which is commonly used in video compression~\cite{wiegand_overview_2003,sullivan_overview_2012} to three-dimensional particle data. A keyframe (analogous to I-frame) transmits the full spatial structure along with absolute quantized coordinates, while a delta frame (analogous to P-frame) transmits only the differences from the preceding frame.
  \end{EN}

\begin{figure*}[t]
    \centering
    \includegraphics[width=1\linewidth]{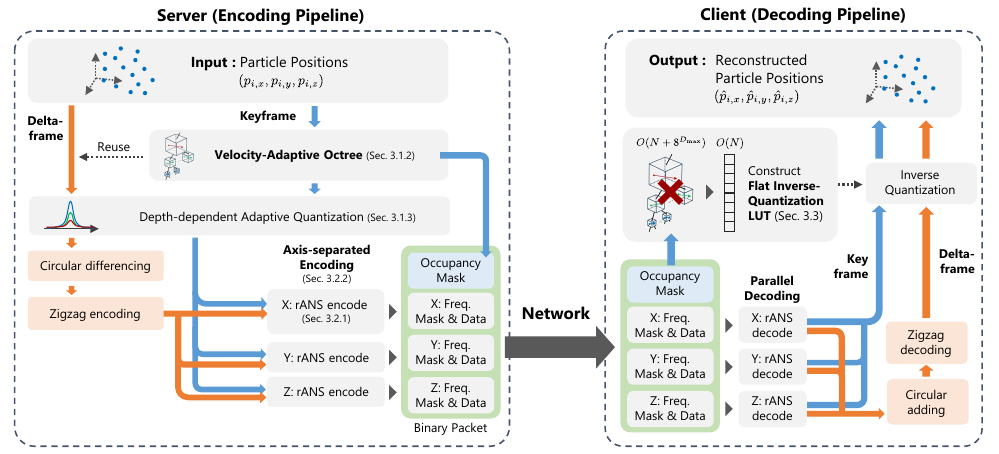}
    \caption{DELUGE encoding and decoding pipeline. Left: the server constructs a velocity-adaptive octree, applies depth-dependent quantization, and compresses with axis-separated rANS. Keyframe paths (blue) carry the full spatial structure including the Occupancy Mask, while delta-frame paths (orange) transmit only ZigZag-coded coordinate differences. Right: the client decodes the three axes in parallel. On keyframe reception, the flat inverse-quantization LUT is built from the octree structure, reducing subsequent inverse quantization. Delta frames reuse the LUT and skip spatial reconstruction.}
    \label{fig:pipeline}
  \end{figure*}

  \begin{JA}
    DELUGEは以下の3段階の符号化パイプラインで構成される（Fig.~\ref{fig:pipeline}）．
    \begin{enumerate}[leftmargin=*]
  \item \textbf{速度適応型ビット配分}（Sec.~\ref{sec:velocity-aware-octree}）：粒子速度に応じてOctree深度を動的に制御し，量子化残差分布をツリー全体で均一化してビット予算を適応的に配分する．
  \item \textbf{軸分離型エントロピー符号化}（Sec.~\ref{sec:adaptive-ans}）：量子化シンボル列を$x$, $y$, $z$軸ごとの独立ストリームとして符号化し，3軸の並列復号により処理を高速化する．
  \item \textbf{平坦逆量子化LUT}（Sec.~\ref{sec:flat-dequant-lut}）：リーフ単位の逆量子化パラメータをパーティクル単位に展開し，木構造探索を排除して$O(N)$アクセスを実現する．
    \end{enumerate}
  \end{JA}
  \begin{EN}
    DELUGE consists of a three-stage encoding pipeline (Fig.~\ref{fig:pipeline}).
    \begin{enumerate}[leftmargin=*]
  \item \textbf{Velocity-adaptive bit allocation} (Sec.~\ref{sec:velocity-aware-octree}): dynamically controls octree depth according to particle velocity, equalizing the quantization residual distribution across the tree and thereby allocating the bit budget adaptively.
  \item \textbf{Axis-separated entropy coding} (Sec.~\ref{sec:adaptive-ans}): encodes quantized symbol sequences as independent streams for the $x$, $y$, and $z$ axes, accelerating processing through parallel decoding of all three axes.
  \item \textbf{Flat inverse-quantization LUT} (Sec.~\ref{sec:flat-dequant-lut}): pre-expands per-leaf inverse-quantization parameters into a per-particle flat array, eliminating tree traversal and enabling $O(N)$ access.
    \end{enumerate}
  \end{EN}

  \begin{JA}
    復号プロセスの全体像はSec.~\ref{sec:decoding-process}で説明し，パケットの詳細は補足資料Sec.~\ref{app:packet-details}に記す．
  \end{JA}
  \begin{EN}
    The overall decoding process is described in Sec.~\ref{sec:decoding-process}, with packet details in supplementary Sec.~\ref{app:packet-details}.
  \end{EN}

  \subsection{Velocity-Adaptive Octree and Depth-Dependent Quantization}\label{sec:velocity-aware-octree}

  \subsubsection{Octree Construction and Particle Reordering}

  \begin{JA}
    流体シミュレーションにおける粒子群は一様ではなく，高速に移動する粒子とほぼ静止している粒子が空間内に混在する．連続フレーム間には強い時間相関が存在し，デルタフレームによる差分符号化の効率は，(i)同一粒子の対応付けが維持されること，(ii)量子化座標の残差分布のエントロピーが低いこと，に強く依存する．
  \end{JA}
  \begin{EN}
    In a fluid simulation, particles are not uniformly distributed; fast-moving particles coexist with nearly stationary ones in the same volume. Consecutive frames exhibit strong temporal correlation, and the efficiency of delta-frame coding depends on two conditions: (i)~particle correspondence between a keyframe and subsequent delta frames is preserved, and (ii)~the entropy of the quantized coordinate residual distribution remains low.
  \end{EN}

  \begin{JA}
    各粒子をOctree~\cite{meagher_geometric_1982}のリーフ内局所座標として量子化する．Octreeを採用するのは，後段のパイプラインが依存する2つの性質のためである．第一に，セルが等方的な立方体であり，その境界がGlobal IDのみから一意に定まる．このためリーフ深度がワールド空間分解能の直接的な指標となり——深度依存量子化（Sec.~\ref{sec:quantization}）が依拠する性質である——リーフごとの幾何情報を伝送する必要もない．KD-treeはこの両方を失う：リーフがデータ依存の非等方な直方体となるため深度と分解能の対応が崩れ，リーフごとの境界の明示的な伝送が必要になり，キーフレームヘッダがリーフ数に比例して肥大する．第二に，二分割で同じリーフ数に達するにはOctreeの約3倍の深度を要し，深度依存のビット配分の下ではリーフあたりのビット予算を圧迫する．予備比較もこれを裏付ける：最良設定のKD-tree変種でもOctree $d{=}3$に対しBD-Rateが$+17.4\%$（$d{=}3$）/$+19.4\%$（$d{=}4$）劣化した（補足資料Sec.~\ref{app:kdtree}）．  \end{JA}
  \begin{EN}
    We quantize each particle's position as local coordinates within an octree leaf~\cite{meagher_geometric_1982}. The octree is chosen for two properties that the rest of the pipeline builds on. First, its cells are isotropic cubes whose bounds are fully determined by the Global ID alone: leaf depth thus becomes a direct proxy for world-space resolution, the property that depth-dependent quantization (Sec.~\ref{sec:quantization}) relies on, and no per-leaf geometry needs to be transmitted. A KD-tree forfeits both: its data-dependent anisotropic leaves break the depth--resolution correspondence and require explicit per-leaf bounds on the wire, inflating keyframe headers in proportion to the leaf count. Second, binary splits need roughly $3\times$ the octree's depth to reach the same leaf count, which erodes the per-leaf bit budget under depth-dependent allocation. A preliminary comparison bears this out: even at its best setting, a KD-tree variant incurred $+17.4\%$ ($d{=}3$) / $+19.4\%$ ($d{=}4$) BD-Rate against Octree $d{=}3$ (supplementary Sec.~\ref{app:kdtree}).  \end{EN}

  \begin{JA}
    本手法では，再帰的にノードを分割して木構造を明示的に構築するのではなく，各パーティクルの座標から所属リーフを直接算出する方式を採用する~\cite{gargantini_linear_1982}．まず全パーティクルの各軸最小値・最大値から立方体バウンディングボックスを構築し，各パーティクルの座標を$[0, 1)$に正規化する．正規化座標を$\hat{p}_{i,a}$（パーティクル$i$の軸$a \in \{x, y, z\}$成分）と表記すると，各レベル$d$におけるオクタント番号は3ビットで表現される．
  \end{JA}
  \begin{EN}
    Rather than constructing the tree through recursive node subdivision, we compute the leaf assignment directly from each particle's coordinates~\cite{gargantini_linear_1982}.  First, we construct a cubic bounding box from the minimum and maximum per-axis values of all particles and normalize each coordinate to $[0, 1)$. Using the normalized coordinate, denoted as $\hat{p}_{i,a}$ for particle $i$ along axis $a \in \{x, y, z\}$, the octant number at each level $d$ is expressed with three bits:
  \end{EN}
  \begin{equation}
    b_{i,d} = b_x + 2\,b_y + 4\,b_z, \quad
    b_a = \lfloor \hat{p}_{i,a} \cdot 2^{d+1} \rfloor \bmod 2
    \label{eq:octant}
  \end{equation}
  \begin{JA}
    ルートからの経路インデックス$k = \sum_{d'=0}^{d_i-1} b_{i,d'} \cdot 8^{d_i - 1 - d'}$を用いて，全ノードを一意に識別する\textbf{Global ID}を定義する．
  \end{JA}
  \begin{EN}
    Using the path index from the root $k = \sum_{d'=0}^{d_i-1} b_{i,d'} \cdot 8^{d_i - 1 - d'}$, we define a \textbf{Global ID} that uniquely identifies every node.
  \end{EN}
  \begin{equation}
    G(d, k) = \frac{8^d - 1}{7} + k
    \label{eq:global-id}
  \end{equation}

  \begin{JA}
    パーティクルはGlobal IDをキーとして安定ソートされ，空間的に近接するパーティクルが配列上で連続するように再配置される．これにより，デルタフレームにおける差分の局所性が高まる．
  \end{JA}
  \begin{EN}
    The particles are then sorted by their Global IDs to ensure that particles in close proximity occupy contiguous positions in the array. This improves the locality of differences in delta frames.
  \end{EN}

  \begin{JA}
    加えて，1個以上のパーティクルを含むリーフノードをアクティブノードと定義し，その数を$N_{\mathrm{leaf}}$とする．各アクティブノードのパーティクル数のプレフィクス和を定義する．
  \end{JA}
  \begin{EN}
    We define a leaf node containing at least one particle as an active node and denote the count of such nodes by $N_{\mathrm{leaf}}$. The prefix sum of the particle counts $n_\ell$ of each active node is defined as $S_0 = 0,\; S_\ell = \sum_{j=1}^{\ell} n_j$.
  \end{EN}

  \subsubsection{Velocity-Adaptive Depth Decision}\label{sec:adaptive-depth}

\begin{figure}[t]
    \centering
    \includegraphics[width=1\linewidth]{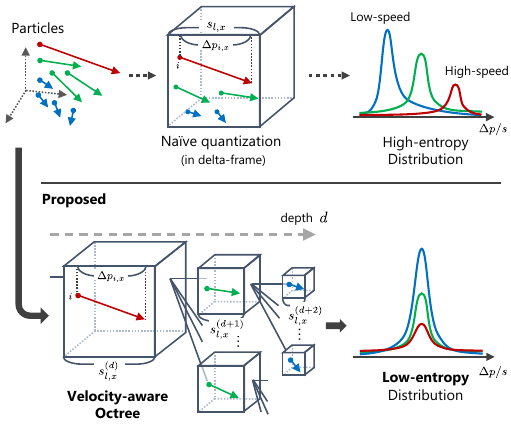}
    \caption{Velocity-adaptive depth assignment and its effect on quantized-residual entropy.}
    \label{fig:velocity-adaptive}
  \end{figure}

  \begin{JA}
    パーティクルの運動速度は一様ではなく，速度の異なる粒子を同一深度のOctreeに配置すると，量子化後の残差分布がノード間で大きく異なることになる．
    この問題に対し，本手法では粒子速度に応じてOctree深度$d_i$を動的に決定する（Fig.~\ref{fig:velocity-adaptive}）．高速粒子には浅いノードを，低速粒子には深いノードを割り当てることで，各粒子の運動量に対するノードサイズの比率をツリー全体で揃える．量子化は各ノードのセルサイズに対して正規化されるため（例：0--255の範囲），ノードサイズが運動量に比例していれば，全粒子の量子化後残差が一定範囲に収まり，残差分布のエントロピーが低減される．
  \end{JA}
  \begin{EN}
    When particles with different velocities are placed at the same octree depth, the post-quantization residual distributions vary widely across nodes.
    To address this, we dynamically determine the octree depth $d_i$ (Fig.~\ref{fig:velocity-adaptive}) ($0 \le d_i \le D_{\max}$, where $D_{\max}$ is the maximum depth) for each particle according to its velocity. Fast particles are assigned to shallow nodes with larger cell sizes, while slow particles are assigned to deep nodes with smaller cell sizes. This maintains a consistent ratio of node size to displacement magnitude across the tree. Since quantization is normalized to each node's cell size (\eg the range 0--255), matching the node size to the displacement magnitude keeps the quantized residuals of all particles within a narrow range, reducing the entropy of the residual distribution.
  \end{EN}

  \begin{JA}
    直近の位置履歴からフレーム間変位スカラーを算出する．
  \end{JA}
  \begin{EN}
    An inter-frame displacement scalar is computed from the recent position history:
  \end{EN}
  \begin{equation}
    \Delta p_i = \bigl\|\mathbf{p}_i^{(t)} - \mathbf{p}_i^{(t-w)}\bigr\|_2
    \label{eq:displacement}
  \end{equation}
  \begin{JA}
    ここで$w$は履歴ウィンドウ幅（本論文では$w = 5$）である．変位$\Delta p_i$を最深レベルのセルサイズ$s = E / 2^{D_{\max}}$（$E$はルートノードの半辺長）で正規化し，2つの閾値$\alpha_1$, $\alpha_2$による段階関数で深度$d_i$を決定する．
  \end{JA}
  \begin{EN}
    where $w$ is the history window width ($w = 5$ in this paper). The displacement $\Delta p_i$ is normalized by the cell size at the deepest level, $s = E / 2^{D_{\max}}$ ($E$ is the half-extent of the root node), and the depth $d_i$ is determined by a step function with two thresholds $\alpha_1$ and $\alpha_2$.
  \end{EN}
  \begin{equation}
    d_i = \begin{cases}
            D_{\max}     & \text{if } \Delta p_i \le \alpha_1 \cdot s \\
            D_{\max} - 1 & \text{if } \alpha_1 \cdot s < \Delta p_i \le \alpha_2 \cdot s \\
            D_{\max} - 2 & \text{if } \Delta p_i > \alpha_2 \cdot s
    \end{cases}
    \label{eq:depth-decision}
  \end{equation}
  \begin{JA}
    ただし$d_i \ge 0$とする．$\alpha_1$, $\alpha_2$は最深セルサイズ$s$に対する変位の比率として定義されるため，シミュレーション空間のスケールに依存しない．実装では$\alpha_1 = 0.05$，$\alpha_2 = 0.10$を採用する．この選択の頑健性は感度分析により検証した：$(\alpha_1, \alpha_2)$を$(0, 0)$（適応無効）から$(0.5, 1.0)$まで11段階でスイープした結果，262K粒子のMLS-MPMシミュレーション（300フレーム）においてbppfの変動は4\%（2.60--2.70），D1 PSNRの変動は0.16\,dB（63.53--63.69\,dB）に留まった．この頑健性は，デルタフレームが全フレームの98.3\%を占め（$\text{kf\_interval} = 60$），深度再割当の影響を受けないことに起因する．
  \end{JA}
  \begin{EN}
    subject to $d_i \ge 0$. Because $\alpha_1$ and $\alpha_2$ are defined as ratios of displacement to the deepest cell size $s$, they are independent of the simulation space scale. In our implementation we set $\alpha_1 = 0.05$ and $\alpha_2 = 0.10$. A sensitivity analysis confirms the robustness of this choice: sweeping 11 configurations from $(\alpha_1, \alpha_2) = (0, 0)$ (adaptive depth disabled) to $(0.5, 1.0)$ on a 262K-particle MLS-MPM simulation (300 frames) yields a bppf range of 2.60--2.70 (4\% variation) and a D1 PSNR range of 63.53--63.69\,dB (0.16\,dB variation). This robustness arises because delta frames, which constitute 98.3\% of all frames at $\text{kf\_interval} = 60$, are unaffected by depth reassignment.
  \end{EN}

  \subsubsection{Depth-Dependent Adaptive Quantization}\label{sec:quantization}

  \begin{JA}
    各パーティクルで決定された深度$\{d_i\}_{i=1}^N$に基づいてOctreeを構築し，各リーフノード$\ell$内のパーティクル座標をリーフの局所座標系で量子化する．リーフの深度$D_L$に応じて量子化ビット数を適応的に決定する．ベース精度を$B$ビット（本論文では$B = 12$）として，
  \end{JA}
  \begin{EN}
    An octree is constructed based on the per-particle depths $\{d_i\}_{i=1}^N$, and the particle coordinates within each leaf node $\ell$ are quantized in the leaf's local coordinate system. The number of quantization bits is determined adaptively according to the leaf depth $D_L$.  The base precision is $B$ bits ($B = 12$ in this paper),
  \end{EN}
  \begin{equation}
    b_\ell = \max(1,\; B - D_L)
    \label{eq:bit-count}
  \end{equation}
  \begin{JA}
    セルサイズ$s_\ell \propto 2^{-D_L}$かつ（$D_L < B$で）$b_\ell = B-D_L$であるため，式~\eqref{eq:bit-count}は大きい浅層セルに多く，小さい深層セルに少ないビットを割り当て，実効ワールド空間量子化ステップ$s_\ell / 2^{b_\ell} \propto 2^{-B}$を深度非依存に保つ．したがって浅いリーフへの割当は量子化を粗くせず，そこに置かれる高速粒子は運動の知覚的マスキングにより瞬時誤差の許容量も大きいと期待できる．深度$D_L$における最大量子化値は$q_{\max} = 2^{b_\ell} - 1$である．  \end{JA}
  \begin{EN}
    Because $s_\ell \propto 2^{-D_L}$ and, for $D_L < B$, $b_\ell = B-D_L$, Eq.~\eqref{eq:bit-count} assigns more bits to larger, shallow cells and fewer to smaller, deep cells, yielding a depth-independent world-space quantization step $s_\ell / 2^{b_\ell} \propto 2^{-B}$. Thus, shallow assignment does not coarsen quantization; fast particles in shallow leaves are also expected to tolerate greater instantaneous error through perceptual masking of motion. The maximum quantization value at depth $D_L$ is $q_{\max} = 2^{b_\ell} - 1$.  \end{EN}

  \begin{JA}
    リーフ$\ell$の境界を$[\mathbf{l}_{\min}, \mathbf{l}_{\max}]$，セルサイズを$\mathbf{s}_\ell = \mathbf{l}_{\max} - \mathbf{l}_{\min}$とする．各軸$a \in \{x, y, z\}$の量子化値は以下で求められる．
  \end{JA}
  \begin{EN}
    Let $[\mathbf{l}_{\min}, \mathbf{l}_{\max}]$ be the bounds of leaf $\ell$ and $\mathbf{s}_\ell = \mathbf{l}_{\max} - \mathbf{l}_{\min}$ its cell size. The quantized value for each axis $a \in \{x, y, z\}$ is computed as
  \end{EN}
  \begin{equation}
    q_{i,a} = \mathrm{clamp}\!\Bigl(
    \Bigl\lfloor
    \frac{p_{i,a} - l_{\min,a}}{s_{\ell,a}} \cdot q_{\max} + 0.5
    \Bigr\rfloor,\; 0,\; q_{\max}
    \Bigr)
    \label{eq:quantize}
  \end{equation}

  \subsection{Axis-Separated Entropy Coding}\label{sec:adaptive-ans}

  \begin{JA}
    前節までの空間分割と量子化により，3次元浮動小数点座標は整数シンボル列$\{s_j\}_{j=1}^{3N}$（$s_j \in [0, 4095]$）に変換される．本節では，これらのシンボル列を圧縮する手法である軸分離型符号化について述べる．
  \end{JA}
  \begin{EN}
    The spatial partitioning and quantization described in the preceding section convert three-dimensional floating-point coordinates into integer symbol sequences $\{s_j\}_{j=1}^{3N}$ ($s_j \in [0, 4095]$). This section describes axis-separated coding, the method used to compress these symbol sequences.
  \end{EN}

  \begin{JA}
    軸分離型符号化の主要な利点は復号の高速化にある．シンボル列を$x$, $y$, $z$軸ごとに独立したストリームに分離することで，3軸の復号を完全に並列化できる．加えて，軸ごとに独立した頻度テーブルを構築するため，各軸の統計的偏りをより正確に捕捉できる．
  \end{JA}
  \begin{EN}
    The main benefit of axis-separated coding is faster decoding. Splitting the symbol sequence into independent streams for the $x$, $y$, and $z$ axes enables decoding of all three axes simultaneously. Additionally, constructing per-axis frequency tables captures axis-specific statistical bias more accurately.
  \end{EN}

  \subsubsection{rANS}

  \begin{JA}
    本手法ではエントロピー符号化方式としてrANS (range variant of Asymmetric Numeral Systems)~\cite{duda2014asymmetricnumeralsystemsentropy}を採用する．rANSは算術符号に匹敵する圧縮率をシンボル単位の償却$O(1)$計算量で実現し，Zstandard~\cite{zstd2016}等の高性能圧縮ライブラリに広く採用されている．
  \end{JA}
  \begin{EN}
    We adopt the range variant of Asymmetric Numeral Systems (rANS)~\cite{duda2014asymmetricnumeralsystemsentropy} as the entropy coding method. rANS achieves compression ratios comparable to arithmetic coding at amortized $O(1)$ cost per symbol and is widely used in modern high-performance compression libraries such as Zstandard~\cite{zstd2016}.
  \end{EN}

  \begin{JA}
    rANSはG-PCCのCABAC~\cite{marpe_context-based_2003}と異なり，シンボル単位で処理が完結するため複数ストリームの並列復号が可能であり，リアルタイムシステムに適している．
  \end{JA}
  \begin{EN}
    Unlike CABAC~\cite{marpe_context-based_2003} in G-PCC, which processes one bit at a time and creates an inherently sequential feedback loop, rANS operates at the symbol level and permits parallel decoding of multiple interleaved streams.
  \end{EN}

  \begin{JA}
    rANSはsemi-static方式~\cite{moffat1995semi_static}で運用する．フレーム全体のシンボル統計からフレームごとの頻度テーブル（$M = 4096$）を構築し，急激な分布変動にも即座に適応する．デルタフレームでは前フレームとの量子化座標差分を循環差分（mod 4096）とZigZag符号化~\cite{protobuf_encoding}により非負整数に変換し，値をゼロ近傍に集中させてエントロピーを低減する．  \end{JA}
  \begin{EN}
    We operate rANS in a semi-static mode~\cite{moffat1995semi_static}: a per-frame frequency table ($M = 4096$) is built from the full symbol statistics, making the coder immediately adaptive to abrupt distribution changes. For delta frames, the quantized coordinate difference from the previous frame is converted to a non-negative integer via circular differencing (mod 4096) and ZigZag encoding~\cite{protobuf_encoding}, concentrating values near zero and reducing entropy.  \end{EN}

  \subsubsection{Axis-Separated Encoding}\label{sec:ans-planar}

  \begin{JA}
    パーティクルの3次元座標は軸ごとに異なる統計的分布を持つ．物理シミュレーションでは重力の影響により$y$軸（鉛直方向）の分布は$x$, $z$軸（水平方向）と顕著に異なる．インターリーブ配置$x_0 y_0 z_0 x_1 \ldots$ではこれらの異なる統計特性が混合され，頻度テーブルが各軸の分布の加重平均となり，圧縮効率が低下する．
  \end{JA}
  \begin{EN}
    Three-dimensional particle coordinates exhibit distinct statistical distributions along each axis. In physics simulations, gravity causes the $y$-axis (vertical) distribution to differ markedly from the $x$ and $z$ (horizontal) distributions. An interleaved layout $x_0 y_0 z_0 x_1 \ldots$ mixes these distinct statistics so that the resulting frequency table becomes a weighted average of the per-axis distributions, reducing compression efficiency.
  \end{EN}

  \begin{JA}
    本手法では，この軸間の統計的独立性を活用するため，semi-static rANSの手順（補足資料Sec.~\ref{app:entropy-details}参照）を各軸のシンボル列$\mathbf{s}^{(a)}$（$a \in \{x, y, z\}$）に独立に適用し，軸ごとの頻度テーブル$\{f_s^{(a)}\}$を構築して軸別のrANS符号化バイトストリームを生成する．
  \end{JA}
  \begin{EN}
    To exploit this inter-axis statistical independence, we apply the semi-static rANS procedure (detailed in the supplementary material, Sec.~\ref{app:entropy-details}) independently to each axis's symbol sequence $\mathbf{s}^{(a)}$ ($a \in \{x, y, z\}$), constructing a per-axis frequency table $\{f_s^{(a)}\}$ and generating axis-specific rANS-coded byte streams.
  \end{EN}
  \begin{equation}
    \{s_j\}_{j=1}^{3N} \xrightarrow{\text{deinterleave}}
    \mathbf{s}^{(x)}, \mathbf{s}^{(y)}, \mathbf{s}^{(z)}
    \quad (|\mathbf{s}^{(a)}| = N)
    \label{eq:planar-deinterleave}
  \end{equation}

  \begin{JA}
    軸分離型rANSの核心的な設計要素は，3軸のペイロードを物理的に独立した連続領域としてパケット内に配置する点にある．各軸のパックされたペイロードを$P_x$, $P_y$, $P_z$と表記すると，エントロピーペイロードは以下の形式をとる．
  \end{JA}
  \begin{EN}
    A central design element of axis-separated rANS is placing the three axes' payloads as physically independent contiguous regions within the packet. Denoting the packed payloads of each axis as $P_x$, $P_y$, $P_z$, the entropy payload takes the following form:
  \end{EN}
  \begin{equation}
    [\,|P_x|\,(4\text{B})\;]\;
    [\,|P_y|\,(4\text{B})\;]\;
  [\,P_x\,]\;[\,P_y\,]\;[\,P_z\,]
    \label{eq:planar-format}
  \end{equation}
  \begin{JA}
    ここで$|P_x|$, $|P_y|$は各ペイロードのバイト長を格納する4バイト（32ビット）符号なし整数のヘッダである．
  \end{JA}
  \begin{EN}
    Here $|P_x|$ and $|P_y|$ are headers storing the byte length of each payload as 4-byte (32-bit) unsigned integers.
  \end{EN}

  \begin{JA}
    この物理分離構造により，復号側では$|P_x|$, $|P_y|$を読み取るだけで3つのペイロードの開始位置が即座に確定し，3軸の復号を完全に並列に実行できる．各ペイロードは自己完結的であり，頻度テーブル，ANS初期状態，および再正規化バイトストリームを含むため，他軸のデータを一切参照せずに独立に復号が完結する．
  \end{JA}
  \begin{EN}
    This physical separation allows the decoder to determine the starting positions of all three payloads by reading only $|P_x|$ and $|P_y|$, enabling fully parallel decoding of the three axes. Each payload is self-contained: it includes the frequency table, the ANS initial state, and the renormalized byte stream, so decoding completes independently without referencing any other axis's data.
  \end{EN}

  \subsection{Flat Inverse-Quantization LUT}\label{sec:flat-dequant-lut}

  \begin{JA}
    式~\eqref{eq:quantize}に対応する逆量子化では，各パーティクルが所属するリーフの境界$l_{\min,a}$，セルサイズ$s_{\ell,a}$，量子化ビット数$b_\ell$が必要となる．素朴な実装はリーフ単位の二重ループとなり，$D_{\max} = 7$では最大$2 \times 10^6$のリーフが存在しうるため分岐予測が困難である．
  \end{JA}
  \begin{EN}
    Inverse quantization corresponding to Eq.~\eqref{eq:quantize} requires each particle's leaf bounds $l_{\min,a}$, cell size $s_{\ell,a}$, and bit count $b_\ell$. A na\"ive nested loop over up to $8^7 \approx 2 \times 10^6$ leaves at $D_{\max} = 7$ makes branch prediction difficult.
  \end{EN}

  \begin{JA}
    この問題を解決するため，キーフレーム受信時にリーフ単位の逆量子化パラメータをパーティクル単位の粒度に展開する平坦逆量子化LUTを構築する．長さ$3N$の浮動小数点配列を2本定義し，パーティクル$i$の軸$a$に対する要素インデックスを$3i + a$とする．オフセット配列$\mathcal{A}$とスケール配列$\mathcal{B}$は以下で与えられる．
  \end{JA}
  \begin{EN}
    To resolve this issue, we create a flat inverse-quantization LUT upon receiving a keyframe. This LUT expands the parameters from leaf-level to particle-level granularity. We define two floating-point arrays of length $3N$, where the element index for particle $i$ along axis $a$ is $3i + a$. The offset array $\mathcal{A}$, and the scale array $\mathcal{B}$ are given by
  \end{EN}
  \begin{align}
    \mathcal{A}[3i + a] &= l_{\min,a}^{(\ell)} \\
    \mathcal{B}[3i + a] &= \frac{s_{\ell,a}}{2^{b_\ell} - 1}
  \end{align}
  \begin{JA}
    ここで$\ell$はパーティクル$i$が属するリーフである．$\mathcal{A}$はリーフ境界の原点座標を，$\mathcal{B}$は1量子化ステップあたりのワールド座標幅を格納する．逆量子化式$\hat{p}_{i,a} = l_{\min,a} + q_{i,a} \cdot s_{\ell,a} / q_{\max}$における除算$s_{\ell,a} / q_{\max}$を事前計算して$\mathcal{B}$に格納することで，復号ループ内の除算を完全に排除する．
  \end{JA}
  \begin{EN}
    where $\ell$ is the leaf to which particle $i$ belongs. $\mathcal{A}$ stores the leaf boundary origin and $\mathcal{B}$ stores the world-coordinate width per quantization step. The division $s_{\ell,a} / q_{\max}$ in the inverse-quantization formula $\hat{p}_{i,a} = l_{\min,a} + q_{i,a} \cdot s_{\ell,a} / q_{\max}$ is precomputed and stored in $\mathcal{B}$, completely eliminating division from the decoding loop.
  \end{EN}

  \subsubsection{Single-Loop Inverse Quantization}

  \begin{JA}
    LUT構築後，逆量子化はリーフ構造を一切参照せず，長さ$3N$の単一ループで完結する．
  \end{JA}
  \begin{EN}
    Once the LUT is built, inverse quantization completes in a single loop of length $3N$ with no reference to the leaf structure.
  \end{EN}
  \begin{equation}
    \hat{p}_{i,a} = \mathcal{A}[3i + a] + q_{i,a} \cdot \mathcal{B}[3i + a]
    \label{eq:flat-dequant}
  \end{equation}
  \begin{JA}
    この構造により，反復回数$3N$固定の単一ループとなり，全分岐が排除されてループ本体は乗算と加算のみで構成される．配列$\mathcal{A}$, $\mathcal{B}$, $q$はパーティクル順に連続配置されるため，キャッシュ効率が高くSIMD自動ベクトル化の理想的な対象となる．
    LUTはキーフレーム受信時（60フレームに1回）に$O(N)$で構築され，追加メモリコストは$24N$バイト（$N = 65{,}536$で約1.6\,MB）である．
  \end{JA}
  \begin{EN}
    This structure yields a fixed-length loop of $3N$ iterations, eliminates all branches, and reduces the loop body to multiply-add operations. The arrays $\mathcal{A}$, $\mathcal{B}$, and $q$ are laid out contiguously in particle order, maximizing cache efficiency and making the loop an ideal candidate for SIMD auto-vectorization.
    The LUT is built in $O(N)$ upon keyframe reception (once every 60 frames); the additional memory cost is $24N$ bytes, which is approximately 1.6\,MB for $N = 65{,}536$.
  \end{EN}

  \begin{JA}
    キーフレームパケットにはバウンディングボックス，Octreeの占有マスク，プレフィクス和（各リーフのパーティクル数の累積和），頻度テーブル，rANSストリームが格納される．デルタフレームでは空間構造を省略し，頻度テーブルとrANSストリームのみを伝送する．
  \end{JA}
  \begin{EN}
    A keyframe packet contains the bounding box, an octree occupancy mask, a prefix sum of per-leaf particle counts, the frequency table, and the rANS stream. Delta frames omit the spatial structure and transmit only the frequency table and the rANS stream, reusing the structure from the most recent keyframe.
  \end{EN}

  \subsection{Decoding Process}\label{sec:decoding-process}
  \label{sec:packet-details}
  \begin{JA}
    クライアント側では，受信パケットから以下の4ステップで座標を復元する．
    \begin{enumerate}[label=\arabic*., leftmargin=*]
  \item 空間構造の再構築（キーフレームのみ）：占有マスクを解析してOctree階層を復元する．各ノードの境界はGlobal IDから一意に決定される．平坦逆量子化LUTもこの時点で構築する．
  \item 統計モデルの復元：周波数マスクと周波数値から頻度テーブルを再構築する．
  \item ANS復号：軸分離rANSにより，3軸のペイロードを並列に復号し，整数シンボル列$\{s_j\}$を得る．各シンボルは$M = 4096$エントリのスロット→シンボル変換テーブルにより$O(1)$で特定される．
  \item 逆量子化：キーフレームでは量子化値を直接逆量子化する．デルタフレームでは，まず差分を$\delta = (z \gg 1) \oplus -(z \mathbin{\&} 1)$で復元し，現フレームの量子化値を$q^{(t)} = (q^{(t-1)} + \delta) \bmod 4096$で得た後，平坦LUTにより$\hat{p}_{i,a} = \mathcal{A}[3i + a] + q_{i,a} \cdot \mathcal{B}[3i + a]$で浮動小数点座標を復元する．
    \end{enumerate}
  \end{JA}
  \begin{EN}
    On the client side, coordinates are reconstructed from the received packet through the following steps.
    \begin{enumerate}[label=\arabic*., leftmargin=*]
  \item Spatial structure reconstruction (keyframe only): the Occupancy Mask is parsed to recover the octree hierarchy. Each node's bounds are uniquely determined from its Global ID. The flat inverse-quantization LUT is built at this point.
  \item Statistical model reconstruction: the frequency table is rebuilt from the Frequency Mask and Frequency Values.
  \item ANS decoding: in axis-separated rANS, the three axes' payloads are decoded in parallel to obtain the integer symbol sequence $\{s_j\}$. Each symbol is identified in $O(1)$ via a slot-to-symbol conversion table of $M = 4096$ entries.
  \item Inverse quantization: for keyframes, quantized values are directly inverse-quantized. For delta frames, the difference is first recovered as $\delta = (z \gg 1) \oplus -(z \mathbin{\&} 1)$, the current frame's quantized value is obtained as $q^{(t)} = (q^{(t-1)} + \delta) \bmod 4096$, and the flat LUT then restores floating-point coordinates via $\hat{p}_{i,a} = \mathcal{A}[3i + a] + q_{i,a} \cdot \mathcal{B}[3i + a]$.
    \end{enumerate}
  \end{EN}

  \section{System Architecture}\label{sec:architecture}

  \begin{JA}
    提案手法の実用性を検証するため，Webブラウザ（WASM+SIMD）およびApple Vision Pro（ネイティブSwift+Rust FFI）の2プラットフォームにエンドツーエンドのストリーミングクライアントを実装した．本システムはシミュレーションを実行するバックエンド，効率的な通信を担うストリーミング層，およびレンダリングを行うクライアントの3層から構成される．
  \end{JA}
  \begin{EN}
    To validate the practicality of the proposed method, we implemented end-to-end streaming clients on two platforms: a web browser (WASM+SIMD) and Apple Vision Pro (native Swift with Rust FFI). The system comprises three layers: a back-end that executes the simulation, a streaming layer that handles efficient communication, and clients that perform rendering.
  \end{EN}

  \subsection{Server and Simulation}

  \begin{JA}
    サーバは2系統で構成される．DELUGE条件ではRustサーバ（axum + tokio）を用い，WebSocket~\cite{fette_websocket_2011}経由で複数クライアントと通信する．TMC13条件ではPythonサーバ（FastAPI/Uvicorn）を用いる．Rustサーバではtokio::time::intervalとドリフト補正（MissedTickBehavior::Skip）による16.6ms周期のタイマー駆動でシミュレーションと符号化を逐次実行し，符号化済みパケットを全クライアントにブロードキャストするため，符号化コストはクライアント数に依存しない．物理エンジンにはMLS-MPM~\cite{hu_moving_2018}を採用し，Metal/CUDAのGPUカーネルをC FFI経由で呼び出すことで$10^4$--$10^5$個規模の粒子を駆動する．
  \end{JA}
  \begin{EN}
    The server comprises two configurations: a Rust server (axum + tokio) for the DELUGE condition and a Python server (FastAPI/Uvicorn) for the the Control condition (TMC13-based), both communicating with multiple clients via WebSocket~\cite{fette_websocket_2011}. The Rust server uses a tokio::time::interval timer with drift correction (MissedTickBehavior::Skip) to drive the 16.6~ms simulation--encoding loop; the encoded packet is broadcast to all clients, making encoding cost independent of client count. The physics engine uses MLS-MPM~\cite{hu_moving_2018} with Metal/CUDA GPU kernels invoked via C FFI, driving $10^4 \sim 10^5$-scale particle simulations.
  \end{EN}

  \subsection{Web Client}\label{sec:web-client}

  \begin{JA}
    Webクライアントでは，DELUGEコーデックのRustコードをWebAssembly (WASM+SIMD) にコンパイルしてデコーダとして使用する．復号処理はWeb Workerで実行され，メインスレッドのブロッキングを防止する．復号された座標はThree.js (WebGPU) でレンダリングされる．このクライアントはプラグイン不要で標準Webブラウザ上で動作する．
  \end{JA}
  \begin{EN}
    The web client compiles the same Rust codec to WebAssembly (WASM+SIMD) for decoding. Decoding runs in a web worker to avoid blocking the main thread, and the reconstructed coordinates are rendered via Three.js (WebGPU). This client requires no plugins and runs in standard web browsers.
  \end{EN}

  \subsection{visionOS Client}\label{sec:visionos-client}

  \begin{JA}
    ユーザ実験（Sec.~\ref{sec:results-perceptual}, \ref{sec:results-collab}）は，Apple Vision Pro上で動作するvisionOSネイティブクライアントを用いて実施した．デコーダはWebクライアントと同一のRustコードベースをstaticライブラリとしてC FFI経由で呼び出す（アルゴリズムは同一，実行環境のみ異なる）． visionOSクライアントではGPUインスタンシングによるicosphereメッシュで粒子を描画する．各粒子位置に対して1段細分割
    のicosphere（42頂点，80三角形）を1インスタンスし，ARKitによるハンドトラッキング（90\,Hz）で流体への直接インタラクションを実現する．マルチユーザ環境では画像マーカーベースの座標整合により空間を統一する．実装詳細は補足資料（Sec.~\ref{app:visionos-details}）に記載する．
  \end{JA}
  \begin{EN}
    The user experiments (Sec.~\ref{sec:results-perceptual}, \ref{sec:results-collab}) were conducted using a native visionOS client on Apple Vision Pro. The decoder invokes the same Rust codebase as the web client, linked as a static library via C FFI (identical algorithms, different execution environment only). The visionOS client renders particles as GPU-instanced icosphere meshes. Each decoded particle position drives
    one instance of a subdivided icosphere (42 vertices, 80 triangles), and ARKit hand tracking at 90\,Hz enables direct fluid interaction. In the multi-user setting, image-marker-based coordinate alignment unifies the simulation space across devices. Implementation details are provided in the supplementary material (Sec.~\ref{app:visionos-details}).
  \end{EN}

  \section{Experiments}\label{sec:experiment}

  \begin{JA}
    本章では，提案手法を4つの観点から評価する．まず実験セットアップを述べ（Sec.~\ref{sec:eval-setup}），次に標準動的点群データセットにおける圧縮性能（Sec.~\ref{sec:results-compression}），各手法の寄与を分離するアブレーション実験（Sec.~\ref{sec:results-ablation}），リアルタイム流体インタラクションにおける知覚品質評価（Sec.~\ref{sec:results-perceptual}），および2人協調タスク評価（Sec.~\ref{sec:results-collab}）を報告する．
  \end{JA}
  \begin{EN}
    This section evaluates the proposed method from four perspectives. We first describe the experimental setup (Sec.~\ref{sec:eval-setup}), then report compression performance on standard dynamic point cloud datasets (Sec.~\ref{sec:results-compression}), ablation experiments isolating the contribution of each technique (Sec.~\ref{sec:results-ablation}), perceptual quality evaluation in real-time fluid interaction (Sec.~\ref{sec:results-perceptual}), and a two-person collaborative task evaluation (Sec.~\ref{sec:results-collab}).
  \end{EN}

  \subsection{Experimental Setup}\label{sec:eval-setup}

  \subsubsection{Datasets}\label{sec:eval-datasets}
  \begin{JA}
    DELUGEの主たる対象である流体シミュレーションデータでの圧縮性能を評価するために，MLS-MPMシミュレーションから生成したデータセットを用いる．

    \paragraph{シミュレーション条件}
    流体シミュレーションにはMLS-MPM (Moving Least Squares Material Point Method)~\cite{hu_moving_2018}をAPIC転送スキーム~\cite{jiang_affine_2015}および二次B-spline重み関数とともに用いる．
    シミュレーション領域は一辺$L$（無次元）の立方体であり，粒子密度を一定に保つため$L = 100 \times (N / 65{,}536)^{1/3}$で設定する（$L{=}100, 126, 159$）．オイラーグリッドは$\min(128, \lfloor L \rfloor)^3$セル（セルサイズ$h = L / \text{gridDim}$；65Kでは$h{=}1.0$，131Kでは$h{=}1.0$，262Kでは$h{=}1.24$）を使用する．

    初期フレームにおいて$N$個の粒子をシミュレーション領域の90\%（壁からのマージン10\%）の範囲に一様ランダム配置し，初速度はゼロとする．
    時間刻みは$\Delta t = 0.006\,\mathrm{s}$，サーバは60\,fpsで駆動する．

    \paragraph{構成モデル}
    流体は弱圧縮性ニュートン流体としてモデル化する．
    圧力はTait状態方程式
    $p = K\bigl[(\rho/\rho_0)^\gamma - 1\bigr]^+$
    （$K{=}2000$，$\rho_0{=}1.0$，$\gamma{=}5$）で計算し，コーシー応力は$\boldsymbol{\sigma} = -p\,\mathbf{I} + 2\mu\,\mathrm{sym}(\mathbf{C})$（$\mu{=}0.01$は動粘性係数，$\mathbf{C}$はAPICアフィン速度行列）とする．重力は$\mathbf{g}=(0,\, {-}9.8,\, 0)\;\mathrm{m/s^2}$である．

    \paragraph{境界条件}
    領域の6面にはグリッド端2セル幅で速度をゼロに固定するno-slip壁を設ける．G2P転送時には予測的壁補正（$k_w{=}0.3$，lookahead $3\Delta t$）を追加適用する．固体障害物はSDF (signed distance field) プリミティブで表現し，表面近傍$1.5h$以内の粒子にペナルティ力を印加，貫通粒子は法線方向に投影・速度反射する．

    \paragraph{データセット構成}
    粒子数$N \in \{65{,}536,\; 131{,}072,\; 262{,}144\}$の3条件（$L{=}100, 126, 159$）について，それぞれ異なるランダムシードで10シーケンス（各600フレーム，10秒）を生成し，合計30シーケンスを得た．$L$を$N$に応じて変動させることで単位体積あたりの粒子密度を一定に保ち，粒子数間のスケーラビリティ比較を公平にしている．RD曲線の評価には65Kおよび262Kの2条件を用い，131Kは処理時間の評価にのみ使用する．
  \end{JA}
  \begin{EN}
    To evaluate compression performance on fluid simulation data, we use datasets generated from MLS-MPM simulations.

    \paragraph{Simulation setup}
    The fluid simulation uses the Moving Least Squares Material Point Method (MLS-MPM)~\cite{hu_moving_2018} with an Affine Particle-In-Cell (APIC) transfer scheme~\cite{jiang_affine_2015} and quadratic B-spline weight functions.
    The simulation domain is a cubic box of side length $L = 100 \times (N / 65{,}536)^{1/3}$ (dimensionless units), yielding $L{=}100, 126, 159$ for the three particle counts so that particle density remains constant.
    An Eulerian background grid of $\min(128, \lfloor L \rfloor)^3$ cells (cell size $h = L / \text{gridDim}$; $h{=}1.0$ for 65K, $h{=}1.0$ for 131K, $h{=}1.24$ for 262K) is used for particle--grid transfers.

    At the initial frame, $N$ particles are placed uniformly at random within 90\% of the domain volume (10\% wall margin), with zero initial velocity.
    The time step is $\Delta t = 0.006\,\mathrm{s}$, and the simulation advances at 60\,fps.

    \paragraph{Constitutive model}
    The fluid is modeled as a weakly compressible Newtonian fluid.
    Pressure is computed via the Tait equation of state:
    $p = K\bigl[(\rho/\rho_0)^\gamma - 1\bigr]^+$,
    with bulk modulus $K{=}2000$, rest density $\rho_0{=}1.0$, and exponent $\gamma{=}5$.
    The Cauchy stress tensor is $\boldsymbol{\sigma} = -p\,\mathbf{I} + 2\mu\,\mathrm{sym}(\mathbf{C})$, where $\mu{=}0.01$ is the dynamic viscosity and $\mathbf{C}$ is the APIC affine velocity matrix.
    Gravity is $\mathbf{g} = (0,\, {-}9.8,\, 0)\;\mathrm{m/s^2}$.

    \paragraph{Boundary conditions}
    No-slip walls are enforced on all six faces of the domain by clamping grid velocities to zero within a 2-cell margin.
    A predictive wall correction (stiffness $k_w{=}0.3$, lookahead $3\Delta t$) is additionally applied during the G2P transfer to prevent boundary penetration.
    Solid obstacles are represented as signed distance field (SDF) primitives; particles within $1.5h$ of a surface receive penalty forces, and penetrating particles are projected outward with velocity reflection along the surface normal.

    \paragraph{Dataset composition}
    We generate 10 sequences of 600 frames (10\,s each) with distinct random seeds for three particle counts $N \in \{65{,}536,\; 131{,}072,\; 262{,}144\}$ ($L{=}100, 126, 159$), yielding 30 sequences in total.
    The domain size $L$ is scaled with $N$ to maintain a constant particle density per unit volume, ensuring fair scalability comparisons across particle counts.
    RD curve evaluation uses the 65K and 262K datasets; the 131K dataset is used only for processing time evaluation.
  \end{EN}

  \subsubsection{Baselines}\label{sec:eval-baselines}
  \begin{JA}
    以下の手法と比較する．
    \begin{itemize}[leftmargin=*]
  \item \textbf{G-PCC} (TMC13 v23.0-rc2)~\cite{gpcc2020}:
  MPEG標準のOctreeベース幾何圧縮．CTCのLossyプロファイルに準拠し，\texttt{positionQuantizationScale} (pqs) $\in \{0.015625, 0.03125, 0.0625, 0.125, 0.25, 0.5, 1.0\}$の7ポイントでRD特性を取得する．
  \item \textbf{Draco}~\cite{draco_github}:
  kd-treeベースの軽量点群圧縮ライブラリ．Web環境での利用を主な対象とする．量子化ビット数 qp $\in \{6, 7, 8, 10, 11, 12, 14\}$の7ポイントでRD特性を取得する．
    \end{itemize}
  \end{JA}
  \begin{EN}
    We compare against the following methods.
    \begin{itemize}[leftmargin=*]
  \item \textbf{G-PCC} (TMC13 v23.0-rc2)~\cite{gpcc2020}:  the MPEG standard octree-based geometry codec. We follow the Common Test Conditions (CTC) Lossy profile and set \texttt{positionQuantizationScale} (pqs) $\in \{0.015625, 0.03125, 0.0625, 0.125, 0.25, 0.5, 1.0\}$ to obtain seven RD operating points.
  \item \textbf{Draco}~\cite{draco_github}:
  a lightweight kd-tree-based point cloud compression library designed primarily for web deployment. RD points are obtained at seven quantization levels qp $\in \{6, 7, 8, 10, 11, 12, 14\}$.
    \end{itemize}
  \end{EN}

  \begin{JA}
    G-PCCおよびDracoは各フレームを独立に符号化する．DELUGEはI/P-frame構造による時間的冗長性の活用を含むため，公平性のためI-frame単独の性能とI+P-frameの総合性能の両方を報告する．
  \end{JA}
  \begin{EN}
    G-PCC and Draco encode each frame independently (intra-frame only). Because DELUGE exploits temporal redundancy through its I/P-frame structure, we report both I-frame-only and combined I+P-frame performance for fairness.
  \end{EN}

  \subsubsection{Evaluation Metrics}\label{sec:eval-metrics}
  \begin{JA}
    \begin{itemize}[leftmargin=*]
  \item \textbf{D1 PSNR}（Point-to-Point PSNR）：幾何品質の客観指標．$\text{D1 PSNR} = 10 \log_{10}(3 \cdot \text{peak}^2 / \text{MSE})$（$\text{peak}$はバウンディングボックス対角長）．MPEG G-PCC CTC準拠のpc\_errorツールで計測する．計算コスト削減のため10フレームおき（600フレーム中60フレーム）に計測する．
  \item \textbf{bits/point/frame (bppf)}：1点1フレームあたりの符号化ビット数．  \item \textbf{BD-Rate}：Bjøntegaard Delta Rate~\cite{schwarz_emerging_2019}．同一D1 PSNRにおけるビットレート差（\%）．G-PCCをanchorとする．  \item \textbf{処理時間}：符号化・復号の所要時間（ms/frame）．DELUGEの復号はネイティブに加えWebAssembly (WASM+SIMD) 上でも計測する．    \end{itemize}
  \end{JA}
  \begin{EN}
    \begin{itemize}[leftmargin=*]
  \item \textbf{D1 PSNR} (Point-to-Point PSNR):
  an objective geometry quality metric, $\text{D1 PSNR} = 10 \log_{10}(3 \cdot \text{peak}^2 / \text{MSE})$, where $\text{peak}$ is the bounding-box diagonal length. Measured with the pc\_error tool following the MPEG G-PCC CTC. To reduce computational cost, D1 PSNR is evaluated on every 10th frame (60 evaluation frames per 600-frame sequence).
  \item \textbf{bits/point/frame (bppf)}: the number of coded bits per point per frame.  \item \textbf{BD-Rate}: Bjøntegaard Delta Rate~\cite{schwarz_emerging_2019}, quantifying the bitrate difference (\%) at equal D1 PSNR. G-PCC serves as the anchor.  \item \textbf{Processing time}: encoding and decoding time (ms/frame). DELUGE decoding is measured natively and additionally on WebAssembly (WASM+SIMD).    \end{itemize}
  \end{EN}

  \subsubsection{Environment}\label{sec:eval-env}
  \begin{JA}
    Table~\ref{tab:eval_environment}に評価環境を示す．圧縮性能とアブレーション実験はDesktop環境で実施し，処理時間の評価ではVRクライアントも含めた2環境で計測する．  \end{JA}
  \begin{EN}
    Table~\ref{tab:eval_environment} lists the evaluation environment. Compression and ablation experiments are conducted on the Desktop configuration; processing time is measured on both the Desktop and VR-client platforms.  \end{EN}

  \begin{table}[t]
    \centering
    \caption{Evaluation environment.}
    \label{tab:eval_environment}
    \scriptsize
    \setlength{\tabcolsep}{2pt}
    \renewcommand{\arraystretch}{0.95}
    \begin{tabular}{@{}l p{0.70\columnwidth}@{}}
      \toprule
      \textbf{Role} & \textbf{Configuration} \\
      \midrule
      Server / Desktop & Apple M2 Max (12-core CPU, 38-core GPU), 64\,GB RAM \\
      VR client & Apple Vision Pro (visionOS 2.2) \\
      \bottomrule
    \end{tabular}
  \end{table}

  \begin{figure}[t]
    \centering
    \includegraphics[width=\linewidth]{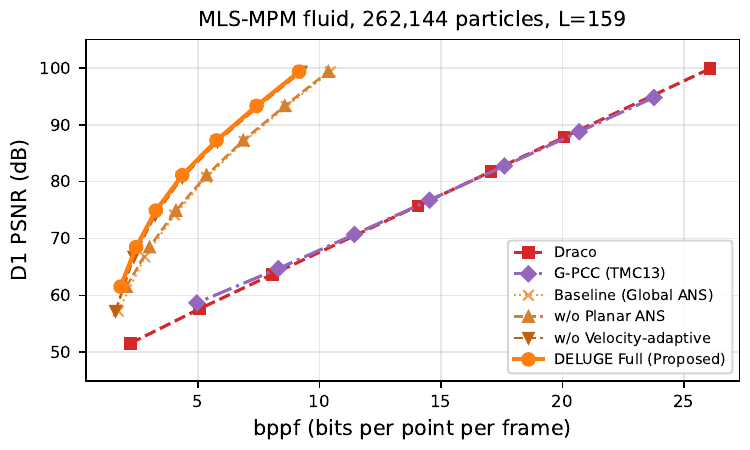}
    \caption{Rate--distortion curves on MLS-MPM fluid simulation at 262K particles ($L{=}159$; 10-sequence average over all 600 frames; D1 PSNR every 10th frame; $D_{\max}{=}7$, keyframe interval 60). The corresponding 65K curves are shown in Supplementary Fig.~\ref{fig:rd-65k}.}
    \label{fig:ablation-rd}
    \vspace{-2mm}
  \end{figure}

  \subsection{Compression Performance}\label{sec:results-compression}

  \subsubsection{Rate--Distortion Comparison}\label{sec:results-rd}
  \begin{JA}
    Table~\ref{tab:rd-comparison}にMLS-MPM流体シミュレーションデータ（262{,}144粒子，10シーケンス平均）におけるDELUGEとG-PCCのRD性能を示す．DELUGEは$D_{\max}=7$，キーフレーム間隔60で，$B$（root\_bits）を8--14に変化させた7レートポイントを報告する．

    Fig.~\ref{fig:ablation-rd}のRDカーブから，3つの特徴が読み取れる．第一に，DELUGEは評価ビットレート全域（1.8--9.2\,bppf）でG-PCCと同等以上のPSNRを達成しながら，高ビットレート域では${\sim}100$\,dB（実質ロスレス）に到達する．    G-PCCは評価レンジ内では${\sim}95$\,dB程度に留まるのに対し，DELUGEはI/P-frame構造と粒子IDベースのデルタ符号化により，ビット増加に対するPSNR改善が高ビットレート域まで持続する．第二に，Dracoは低ビットレート域ではDELUGE・G-PCCの両者に劣るが，高ビットレート域では${\sim}100$\,dBまで到達する．第三に，I/P-frame構造の効果はTable~\ref{tab:rd-comparison}のI/P bppf列に現れる．P-frame bppfはI-frame bppfの15--30\%に圧縮されており，粒子IDが保存されるシミュレーションデータにおいてデルタ符号化が有効に機能している．
    Fig.~\ref{fig:ablation-rd}に262K，Supplementary Fig.~\ref{fig:rd-65k}に65K粒子のRDカーブを示す．両粒子数で同じ傾向がみられ，262KでRD効率が改善するのはoctreeヘッダの固定コストがより多くの粒子に按分されるためである．
  \end{JA}
  \begin{EN}
    Table~\ref{tab:rd-comparison} reports the rate--distortion performance on MLS-MPM fluid simulation data (262{,}144 particles, 10-sequence average). DELUGE uses $D_{\max}{=}7$ with a keyframe interval of 60, sweeping root bits $B$ from 8 to 14 to produce seven rate points.

    Three observations emerge from the RD curves in Fig.~\ref{fig:ablation-rd}. First, DELUGE matches or exceeds G-PCC in PSNR across its entire evaluated bitrate range (1.8--9.2\,bppf) and reaches ${\sim}100$\,dB (near-lossless) at higher rates, whereas G-PCC stays around ${\sim}95$\,dB within the evaluated range.     The I/P-frame structure and particle-ID-based delta coding allow DELUGE to sustain PSNR gains further into the high-bitrate regime. Second, Draco underperforms both DELUGE and G-PCC at low bitrates but reaches ${\sim}100$\,dB at higher rates. Third, the benefit of the I/P-frame structure is evident from the per-type bppf columns in Table~\ref{tab:rd-comparison}: P-frame bppf is 15--30\% of the corresponding I-frame bppf, confirming that delta coding is highly effective for simulation data in which particle IDs are preserved across frames.
    Fig.~\ref{fig:ablation-rd} shows the 262K-particle RD curves, with the 65K result in Supplementary Fig.~\ref{fig:rd-65k}. Both particle counts show the same trends; RD efficiency improves at 262K because the fixed octree header cost is amortized over more particles.
  \end{EN}

  \begin{table}[t]
    \centering
    \caption{DELUGE rate--distortion on MLS-MPM fluid simulation (262{,}144 particles, 10-sequence average, 600 frames each, $D_{\max}{=}7$, keyframe interval 60). BD-Rate (Bjøntegaard with PCHIP interpolation) is reported for the full system (I+P) and for the intra-only I-frame representation; G-PCC/Draco RD curves are shown in Fig.~\ref{fig:ablation-rd}. At 65K particles, the I+P BD-Rate is $-78.4\%$ vs.\ G-PCC and $-76.7\%$ vs.\ Draco.    }
    \label{tab:rd-comparison}
    \scriptsize
    \setlength{\tabcolsep}{3pt}
    \renewcommand{\arraystretch}{0.95}
    \begin{tabular}{@{}l r r r r@{}}
      \toprule
      \textbf{Config} & \textbf{bppf} & \textbf{I bppf} & \textbf{P bppf} & \textbf{D1 PSNR (dB)} \\
      \midrule
      $B{=}8$  & 1.82 & 11.26 & 1.66 & 61.53 \\
      $B{=}9$  & 2.46 & 14.58 & 2.25 & 68.49 \\
      $B{=}10$ & 3.27 & 17.61 & 3.03 & 74.91 \\      $B{=}11$ & 4.36 & 20.60 & 4.08 & 81.13 \\
      $B{=}12$ & 5.77 & 23.55 & 5.47 & 87.24 \\
      $B{=}13$ & 7.41 & 26.43 & 7.09 & 93.30 \\
      $B{=}14$ & 9.17 & 29.43 & 8.82 & 99.35 \\
      \midrule
      \multicolumn{5}{@{}l}{\textbf{BD-Rate vs.\ G-PCC} (full I+P): $-73.5\%$\quad(I only: $+29.5\%$)}\\
      \multicolumn{5}{@{}l}{\textbf{BD-Rate vs.\ Draco} (full I+P): $-72.8\%$\quad(I only: $+27.0\%$)}\\
      \bottomrule
    \end{tabular}
  \end{table}

  \subsubsection{Processing Time}\label{sec:results-speed}
  \begin{JA}
    リアルタイムストリーミングでは符号化・復号処理が60\,fpsの時間制約（16.67\,ms/frame）を満たすことが必須である．
    Table~\ref{tab:processing_time}に，MLS-MPMシミュレーション（65K，131K，262K粒子）におけるDELUGE，G-PCC，Dracoの符号化・復号時間を示す．全コーデックをネイティブin-process計測（フレームごとに1回の符号化/復号呼び出し，subprocessなし）で統一し，公平な比較を行う．全コーデックともTable~\ref{tab:rd-comparison}と同一のデータセット（10シーケンス$\times$全600フレーム）を同一マシン・同一プロトコルで処理し，フレームごとの所要時間の平均値を報告する（DELUGEはI/Pフレームタイプ別に分離）．    G-PCCは\texttt{pqs=0.25}，Dracoは\texttt{qp=8}とし，DELUGEとPSNRが近い条件に揃えた（Sec.~\ref{sec:eval-baselines}）．G-PCC FFIは深い八分木再帰のため64\,MBスタックの専用スレッドで実行した．65K粒子（$L{=}100$）はユーザ実験（Sec.~\ref{sec:results-perceptual}）の構成に，131K・262K粒子（$L{=}126, 159$）はRD評価の構成に対応する．Vision Pro列は，配備クライアントのin-process FFIデコーダを用い，Desktopと同一のデータセット・集計（65K粒子，10シーケンス$\times$600フレーム）で実機計測した．  \end{JA}
  \begin{EN}
    Real-time streaming requires the encoding and decoding processes to complete within the 60\,fps budget of 16.67\,ms per frame. Table~\ref{tab:processing_time} lists the processing times of DELUGE, G-PCC, and Draco on MLS-MPM simulations at 65K, 131K, and 262K particles. All codecs are measured in-process through native harnesses (one encode/decode call per frame, no subprocess overhead) to ensure a fair comparison. All codecs process the same dataset as Table~\ref{tab:rd-comparison} (10 sequences $\times$ 600 frames) on the same machine under an identical protocol, and we report per-frame means (separated by I/P frame type for DELUGE).     G-PCC uses \texttt{pqs\,=\,0.25} and Draco uses \texttt{qp\,=\,8}, chosen to approximately match DELUGE's PSNR operating point (Sec.~\ref{sec:eval-baselines}). The G-PCC FFI runs on a dedicated thread with a 64\,MB stack to accommodate deep octree recursion. The 65K configuration ($L{=}100$) corresponds to the user study setup (Sec.~\ref{sec:results-perceptual}); the 131K and 262K configurations ($L{=}126, 159$) correspond to the RD evaluation. The Vision Pro column is measured on-device through the deployed client's in-process FFI decoders, on the same dataset and aggregation as Desktop (65K particles, 10 sequences $\times$ 600 frames).  \end{EN}

  \begin{table}[t]
    \centering
    \caption{Processing time per frame (ms) on MLS-MPM fluid simulation (DELUGE at $D_{\max}{=}7$, $B{=}12$, ANS Planar, $\text{kf\_interval}{=}60$; measurement conditions are described in the text).}    \label{tab:processing_time}
    \scriptsize
    \setlength{\tabcolsep}{2pt}
    \renewcommand{\arraystretch}{0.95}
    \begin{tabular}{@{}l l r r r r@{}}
      \toprule
      & & \multicolumn{3}{c}{\textbf{Desktop}} & \textbf{Vision Pro} \\
      \cmidrule(lr){3-5} \cmidrule(lr){6-6}
      \textbf{Method} & \textbf{Process} & \textbf{65K} & \textbf{131K} & \textbf{262K} & \textbf{65K} \\
      \midrule
      G-PCC & Encode & 46.6 & 96.4 & 204.9 & --- \\
      G-PCC & Decode & 28.2 & 57.4 & 117.6 & 31.0 \\
      Draco & Encode & 9.8 & 20.0 & 41.0 & --- \\
      Draco & Decode & 3.4 & 6.8 & 13.6 & --- \\
      \midrule
      DELUGE & Encode (I) & 4.4 & 8.5 & 16.2 & --- \\
      DELUGE & Encode (P) & 1.7 & 3.3 & 6.6 & --- \\
      DELUGE & Decode (I) & 2.3 & 4.3 & 8.1 & 2.5 \\
      DELUGE & Decode (P) & 1.4 & 2.8 & 5.5 & 1.6 \\
      DELUGE & Decode (I, WASM) & 2.7 & 5.0 & 9.4 & --- \\
      DELUGE & Decode (P, WASM) & 1.7 & 3.3 & 6.6 & --- \\      \bottomrule
    \end{tabular}
    \vspace{-3mm}
  \end{table}

  \begin{JA}
    DELUGEの復号処理はテーブル参照とビット演算のみで完結するrANSに基づくため，WebAssembly上で高速に動作する：Table~\ref{tab:processing_time}のWASM行は，同一のデコーダをWebAssembly (WASM+SIMD) にコンパイルして同一マシン・同一データで計測したもので（復号出力はネイティブとビット一致），所要時間はネイティブの約1.2倍に留まる．    特にデルタフレームでは空間構造の再構築が不要であり，ANS復号とZigZag逆変換のみで座標を復元できる．
  \end{JA}
  \begin{EN}
    Because DELUGE decoding relies entirely on table lookups and bitwise operations inherent to rANS, it runs efficiently in WebAssembly: the WASM rows of Table~\ref{tab:processing_time} run the identical decoder compiled to WebAssembly (WASM+SIMD) on the same machine and data (decoded output bit-identical to native), at about $1.2\times$ the native decoding time. Delta frames are particularly fast: the spatial structure need not be rebuilt, so coordinate reconstruction reduces to ANS decoding followed by ZigZag inversion.  \end{EN}

  \subsubsection{Preliminary Evaluation on Captured Point Clouds}\label{sec:results-captured}
  \begin{JA}
    DELUGEはシミュレータが付与する粒子IDを前提とするため，フレーム間対応を持たない実計測点群ではI-frame経路のみが利用できる．この適用限界を確認するため，8iVSLF Thaidancer~\cite{krivokuca_8ivslf_2018}およびOwlii basketball\_player/dancer~\cite{xu_owlii_2017}（3シーケンス平均，各300フレーム）で予備評価を行った．復号経路の優位性はそのまま転移する：2.6--3.2M点の3シーケンスにおいてI-frame復号はDELUGEが47--57\,ms vs G-PCC 1{,}019--1{,}166\,ms（各300フレームの平均）で約21倍高速（シーケンス別19.9--21.5倍）であり，シミュレーションデータでの約20倍と整合する（シーケンス別の内訳と詳細な計測条件は補足資料Sec.~\ref{app:captured-details}）．一方，I-frameのみの運用となるためデルタ符号化の利得は消失し（平均bppf 15.46 @ 60.12\,dB），レートではintra専用ベースラインに劣後する．これは実計測データへの拡張のボトルネックが圧縮アルゴリズムではなくID割当てにあることを示している．  \end{JA}
  \begin{EN}
    Because DELUGE presumes simulator-provided particle IDs, captured point clouds without frame-to-frame correspondence can exercise only its I-frame path. To probe this boundary, we ran a preliminary evaluation on 8iVSLF Thaidancer~\cite{krivokuca_8ivslf_2018} and Owlii basketball\_player/dancer~\cite{xu_owlii_2017} (3-sequence mean, 300 frames each). The decode-path advantage transfers directly: across the three 2.6--3.2M-point sequences, DELUGE decodes I-frames in 47--57\,ms versus 1{,}019--1{,}166\,ms for G-PCC (per-frame means over the same 300 frames), ${\sim}21\times$ faster (19.9--21.5$\times$ per sequence), matching the ${\sim}20\times$ advantage on simulation data (per-sequence breakdown and detailed setup in supplementary Sec.~\ref{app:captured-details}). The delta-coding gains, however, disappear because only I-frames can be used (average bppf 15.46 at 60.12\,dB), leaving DELUGE behind the intra-only baselines in rate. This identifies ID assignment---not the compression algorithm---as the bottleneck for extending DELUGE to captured data.
  \end{EN}

  \subsection{Ablation Study}\label{sec:results-ablation}

  \begin{JA}
    提案手法の3つの核心技術——速度適応ビット配分，軸分離型ANS，平坦逆量子化LUT——の寄与を分離するため，MLS-MPM流体シミュレーションデータを用いたアブレーション実験を行う（Fig.~\ref{fig:ablation-rd}）．各技術を個別に除去した変種と，全技術を除去したベースライン（Global ANS）を比較する．
  \end{JA}
  \begin{EN}
    To isolate the contribution of each core technique, velocity-adaptive bit allocation, axis-separated ANS, and the flat inverse-quantization LUT, we conduct an ablation study on the MLS-MPM fluid simulation data (Fig.~\ref{fig:ablation-rd}). 
    we evaluate the full $2^3$ product of the three techniques.
  \end{EN}

  \begin{table}[t]
    \centering
    \caption{Ablation over the full Cartesian product of velocity-adaptive depth (V), axis-separated Planar ANS (P), and the flat inverse-quantization LUT (L) on MLS-MPM fluid simulation (262{,}144 particles, $D_{\max}{=}7$, $B{=}12$, 10 sequences, all 600 frames). bppf/PSNR and Enc./Dec.\ are measured on the same packet streams under the measurement protocol of Table~\ref{tab:processing_time}.}    \label{tab:ablation}
    \scriptsize
    \setlength{\tabcolsep}{2pt}
    \renewcommand{\arraystretch}{0.95}
    \begin{tabular}{@{}c c c r r r r r r r r r r@{}}
      \toprule
      & & & \multicolumn{3}{c}{\textbf{bppf}} & \textbf{PSNR} & \multicolumn{3}{c}{\textbf{Enc.\ (ms)}} & \multicolumn{3}{c}{\textbf{Dec.\ (ms)}} \\
      \cmidrule(lr){4-6} \cmidrule(lr){8-10} \cmidrule(lr){11-13}
      \textbf{V} & \textbf{P} & \textbf{L} & \textbf{avg} & \textbf{I} & \textbf{P} & \textbf{(dB)} & \textbf{avg} & \textbf{I} & \textbf{P} & \textbf{avg} & \textbf{I} & \textbf{P} \\
      \midrule
      Y & Y & Y & 5.77 & 23.55 & 5.47 & 87.24 & 6.7 & 16.2 & 6.6 & 5.6 & 8.1 & 5.5 \\
      Y & Y & N & 5.77 & 23.55 & 5.47 & 87.24 & 6.7 & 16.2 & 6.6 & 6.1 & 8.1 & 6.0 \\
      Y & N & Y & 6.87 & 23.52 & 6.58 & 87.24 & 7.0 & 16.2 & 6.9 & 5.9 & 7.9 & 5.9 \\
      Y & N & N & 6.87 & 23.52 & 6.58 & 87.24 & 7.0 & 16.2 & 6.8 & 6.4 & 7.9 & 6.4 \\
      N & Y & Y & 5.81 & 29.06 & 5.41 & 87.06 & 7.3 & 21.0 & 7.0 & 5.7 & 13.1 & 5.5 \\
      N & Y & N & 5.81 & 29.06 & 5.41 & 87.06 & 7.3 & 21.0 & 7.0 & 7.4 & 13.9 & 7.3 \\
      N & N & Y & 6.91 & 29.02 & 6.53 & 87.06 & 7.6 & 21.0 & 7.3 & 5.9 & 12.8 & 5.8 \\
      N & N & N & 6.91 & 29.02 & 6.53 & 87.06 & 7.6 & 20.9 & 7.3 & 7.8 & 13.7 & 7.7 \\
      \bottomrule
    \end{tabular}
    \vspace{-2mm}
  \end{table}

  \begin{JA}
    各技術の寄与は以下の通りである（Table~\ref{tab:ablation}，RDカーブ全体はFig.~\ref{fig:ablation-rd}参照）．速度適応型深度決定は，高速パーティクルを浅いノードに格上げして量子化残差分布を均一化し，同一ベース量子化（$B{=}12$）でD1 PSNRを0.18\,dB改善するとともに，キーフレームbppfを19.0\%削減し（29.06→23.55），I-frameの符号化・復号を短縮する（21.0→16.2 / 13.1→8.1\,ms）．RDカーブ上では，同一bppfでの利得は最低レート（$B{=}8$）で$+1.2$--$1.5$\,dB，$B{\ge}12$では${\sim}{+}0.3$\,dBに留まる（w/o VelocityをアンカーとしたBD-PSNR $+0.5$\,dB / BD-Rate $-2\%$；Fig.~\ref{fig:ablation-rd}）．したがって，主たる便益は平均RD効率ではなく，キーフレームサイズとI-frameレイテンシの削減である．
    軸分離型ANSは，軸固有の分布を個別に捕捉することで同一PSNRのまま平均bppfを16.0\%削減し（6.87→5.77），3軸の並列復号を可能にする．両者のbppf/PSNRへの相互作用は無視できる大きさであり，寄与は分離可能である．処理時間の観点では，全フレームの98.3\%を占めるP-frame（kf\_interval$=$60）は全8変種で60\,fps予算（16.67\,ms）内に収まり（最悪でもdec(P) 7.7\,ms，enc(P) 6.6--7.3\,ms），I-frameは60フレームに1回のため，そのコスト増（最大enc 21.0 / dec 13.9\,ms）はフレーム平均に影響しない．平坦逆量子化LUTは圧縮率に影響せず，木構造探索を排除して逆量子化工程単体を約8倍高速化する（262K粒子・$D_{\max}{=}7$）．end-to-endでもLUTの無効化はP-frame復号を全構成で一貫して遅くし，その影響はFull構成の9\%（5.5→6.0\,ms）から速度適応を無効化した構成の${\sim}32$\%（5.5→7.3 / 5.8→7.7\,ms）に及ぶ．
  \end{JA}
  \begin{EN}
    From Table~\ref{tab:ablation} and Fig.~\ref{fig:ablation-rd}, velocity-adaptive depth assignment promotes fast-moving particles to shallower nodes, equalizing the quantization residual distribution: at the same base quantization ($B{=}12$) it improves D1 PSNR by 0.18\,dB, cuts keyframe bppf by 19.0\% (29.06$\to$23.55), and shortens I-frame encoding/decoding (21.0$\to$16.2 / 13.1$\to$8.1\,ms). On the RD curves, its matched-bppf gain is $+1.2$--$1.5$\,dB at the lowest rate ($B{=}8$) and shrinks to ${\sim}{+}0.3$\,dB for $B \ge 12$ (BD-PSNR $+0.5$\,dB / BD-Rate $-2\%$ against the w/o-Velocity anchor; Fig.~\ref{fig:ablation-rd}). Thus, its primary benefits are reductions in keyframe size and I-frame latency rather than average RD efficiency.
    Axis-separated ANS captures the per-axis distributions individually, cutting average bppf by 16.0\% at equal PSNR (6.87$\to$5.77) while enabling three-way parallel decoding. The bppf/PSNR interactions between the two are negligible, so their contributions are separable. In terms of processing time, P-frames (98.3\% of all frames at $\text{kf\_interval}{=}60$) fit the 60\,fps budget (16.67\,ms) in all eight variants (worst case dec(P) 7.7\,ms; enc(P) stays at 6.6--7.3\,ms), and I-frames occur only once every 60 frames, so their higher cost (up to 21.0\,ms enc / 13.9\,ms dec) does not affect the per-frame average. The flat inverse-quantization LUT does not affect compression ratio; it accelerates the dequantization step alone by about $8\times$ by eliminating tree traversal (262K particles, $D_{\max}{=}7$). End-to-end, disabling the LUT consistently lengthens P-frame decoding in every configuration, from 9\% in the Full configuration (5.5$\to$6.0\,ms) up to ${\sim}$32\% in the velocity-off configurations (5.5$\to$7.3 / 5.8$\to$7.7\,ms).
  \end{EN}

  \begin{figure*}[t]
    \centering
    \includegraphics[width=\linewidth]{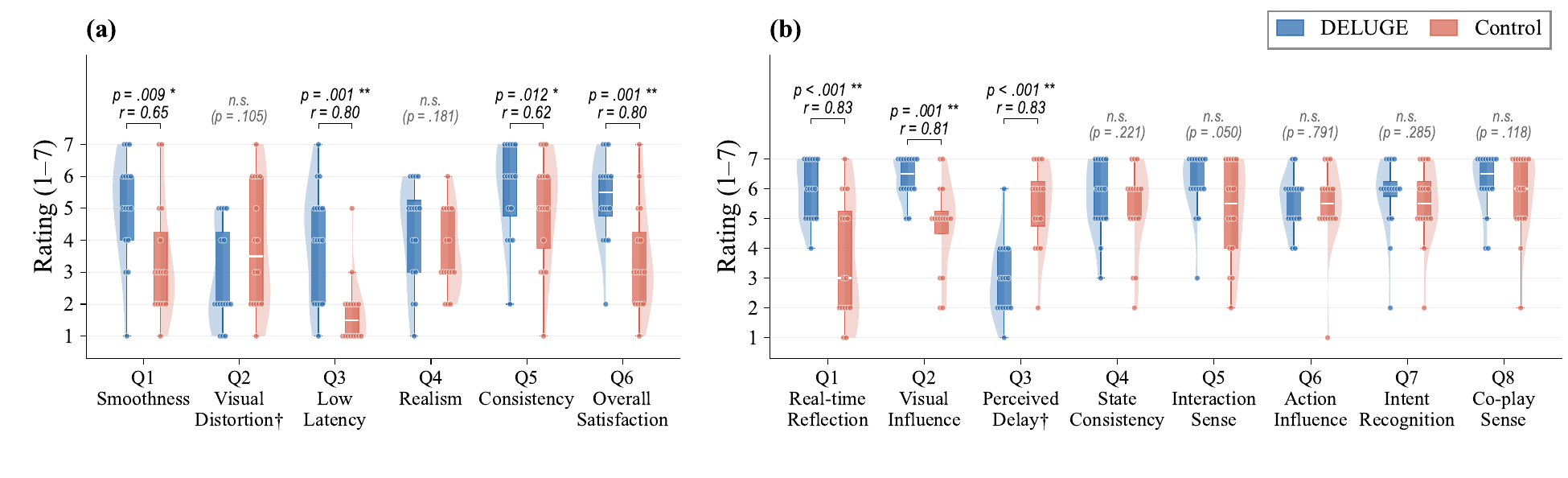}
    \caption{User-study ratings for DELUGE and the TMC13-based Control ($N = 16$, within-participants; Wilcoxon signed-rank tests with Holm correction): (a)~perceptual quality and (b)~collaborative task. Arrows ($\uparrow$/$\downarrow$) indicate the favorable direction.}
    \label{fig:user-study}
  \end{figure*}

  \subsection{Perceptual Quality Evaluation}\label{sec:results-perceptual}

  \subsubsection{Experimental Design}

  \begin{JA}
    圧縮に伴う品質低下がリアルタイム流体インタラクションの体験にどの程度影響するかを評価するため，2条件被験者内比較の主観評価実験を実施した（Cluster, Inc. Research Ethics Committee承認，承認番号2025-015）．事前の検出力分析により，2条件の被験者内計画（対応のある$t$検定，Cohen's $d_z$を効果量指標，$\alpha = .05$，両側，検出力$= .80$）で必要な最小サンプルサイズを$N = 19$と算出し，これを目標として参加者を募集した．実際の分析では順序尺度データに対しWilcoxon符号順位検定を使用するため（Sec.~\ref{sec:results-perceptual-results}），この$t$検定ベースの見積もりは必要サンプルサイズの保守的な下限として機能する．
    22名が実験に参加し，6名を除外した（メガネとヘッドセットの非互換: 3名，ソフトウェア不具合: 2名，通信障害: 1名）．有効な分析対象は16名（男性14名，女性2名，年齢 $M = 30.3$, $SD = 8.2$, 範囲21--48歳），全員が正常視力（矯正含む）を有する．参加者には謝礼として約13\,USDのギフトカードを支給した．実験は3会場・3日間にわたり実施された．22名から16名への減少により，目標効果量（$d_z = 0.68$）に対する達成検出力は約.72に低下した．しかし，有意であった項目の観測効果量（$r = .62$--$.80$）はこの閾値を大幅に上回っており，また非有意であった項目（Q2, Q4）の効果量は小さかった（$r = .33$--$.41$）ことから，非有意の結果は検出力不足ではなく効果自体が小さいことを反映していると考えられる．
  \end{JA}
  \begin{EN}
    To assess how compression-induced degradation affects the experience of real-time fluid interaction, we conducted a within-participants subjective quality experiment comparing two streaming system conditions. The study was approved by the Cluster, Inc. Research Ethics Committee (approval No. 2025-015). An a priori power analysis for a two-condition within-subjects design (paired $t$-test, Cohen's $d_z$, $\alpha = .05$, two-tailed, power $= .80$) indicated a minimum sample size of $N = 19$. Because parametric assumptions may not hold for ordinal Likert data, we use the Wilcoxon signed-rank test for the actual analysis (Sec.~\ref{sec:results-perceptual-results}); the $t$-test-based estimate serves as a conservative lower bound on the required sample size.

    We recruited 22 participants; six were excluded (eyeglasses incompatible with headset: 3, software bugs: 2, connectivity: 1). The final valid sample comprised 16 participants (14 male, 2 female; age $M = 30.3$, $SD = 8.2$, range 21--48), all with normal or corrected-to-normal vision. Each participant received a gift card worth approximately 13\,USD. The experiment was conducted across three venues over three days. This reduction from 22 to 16 lowers the achieved power to approximately .72 for the target effect size ($d_z = 0.68$). However, the observed effect sizes for significant items ($r = .62$--.80) substantially exceed this threshold, and non-significant items (Q2, Q4) showed small effect sizes ($r = .33$--.41), suggesting that the null results reflect genuinely small effects rather than insufficient power.
  \end{EN}

  \begin{JA}
    独立変数はストリーミングシステム条件の2水準であり，DELUGE条件（$D_{\max}{=}7$, $B{=}12$，Rust FFI直接呼出し）とControl条件（TMC13ベースのストリーミングシステム）を比較した．条件の提示順序はカウンターバランスした．各条件で参加者はApple Vision Proを装着し，ハンドトラッキングにより65,536粒子のリアルタイム流体シミュレーション（$\Delta t = 0.003\,\mathrm{s}$，WebSocket経由，Sec.~\ref{sec:visionos-client}）に素手で自由にインタラクションした．体験後にQ1: 滑らかさ，Q2: 視覚的乱れ（逆転項目），Q3: 低遅延，Q4: リアルさ，Q5: 一貫性，Q6: 全体満足度の6項目を7段階リッカート尺度で評価した（詳細な質問文は補足資料Sec.~\ref{app:questionnaire-details}参照）．

    なお，Control条件ではTMC13リファレンスソフトウェアをsubprocess経由で呼び出している．プロセス起動（fork/exec）自体のコストは小さく，同一条件（65{,}536粒子）でsubprocess方式とC FFI in-process方式のフレーム間隔（IFR）を比較すると差は8.1\,ms（IFR中央値69.5\,ms vs 61.4\,ms）である．しかしこの起動コストは，配備パイプライン全体のオーバーヘッドを説明するものではない：ユーザ実験条件には，ラッパーのファイルI/O・シリアライズ，ネットワーク伝送，およびシステム統合のコストが含まれる．実験中に計測された段階別中央値はControl条件でencode 640.8\,ms / network 335.6\,ms / decode 75.3\,ms / E2E 1065.1\,ms（DELUGE条件のE2E中央値は71.5\,ms）であるのに対し，コーデック単体のin-process FFI計測（Table~\ref{tab:processing_time}）ではTMC13のencodeは46.6\,msである．また同セッション中にVision Pro実機で記録された復号の中央値はDELUGE 8.4\,ms / TMC13 80.0\,msであった（study時点のクライアント実装による値；現行実装のコーデック単体実機ベンチはTable~\ref{tab:processing_time}のVision Pro列を参照）．またnetwork中央値の差は，主にペイロードサイズの差（TMC13約891\,KB/frame vs DELUGE約48\,KB/frame）すなわちコーデックのビットレート差（Table~\ref{tab:rd-comparison}）に起因する．したがって本ユーザ実験は，2つの\emph{配備された}インタラクション条件の比較——E2Eレイテンシーが知覚・協調体験に与える影響の検証——として解釈されるべきであり，コーデックカーネル単体の比較ではない．後者はTable~\ref{tab:processing_time}のFFI計測が担う．  \end{JA}
  \begin{EN}
    The independent variable was the streaming system condition with two levels: the DELUGE system ($D_{\max}{=}7$, $B{=}12$; Rust codec invoked via direct FFI) and the Control system (a TMC13-based streaming pipeline). Condition order was counterbalanced. In each condition, participants wore an Apple Vision Pro and interacted freely with a real-time fluid simulation of 65{,}536 particles ($\Delta t = 0.003\,\mathrm{s}$) streamed via WebSocket (Sec.~\ref{sec:visionos-client}), using hand tracking to touch and manipulate the fluid. They then rated six items on a 7-point Likert scale: Q1~Smoothness, Q2~Visual Distortion (reverse-scored), Q3~Low Latency, Q4~Realism, Q5~Consistency, and Q6~Overall Satisfaction (see Sec.~\ref{app:questionnaire-details} for the full item wording).

    Note that the Control condition invoked the TMC13 reference software as a subprocess. The process-launch (fork/exec) cost itself is small: comparing subprocess and C~FFI in-process invocations under the same conditions (65{,}536 particles) shows a difference of 8.1\,ms in the inter-frame interval (IFR median 69.5\,ms vs.\ 61.4\,ms). This launch cost, however, does not account for the overhead of the full deployed pipeline: the user-study condition additionally includes the wrapper's file I/O and serialization, network transport, and system integration costs. The per-stage medians measured during the experiment were, for the Control condition, encode 640.8\,ms / network 335.6\,ms / decode 75.3\,ms / E2E 1065.1\,ms (DELUGE E2E median: 71.5\,ms), whereas the codec-only in-process FFI measurement (Table~\ref{tab:processing_time}) reports TMC13 encoding at 46.6\,ms. On-device decode medians recorded on Vision Pro during the same sessions were 8.4\,ms for DELUGE and 80.0\,ms for TMC13, reflecting the study-time client implementation; the current implementation's codec-only on-device figures appear in the Vision Pro column of Table~\ref{tab:processing_time}.  \end{EN}

  \subsubsection{Results}\label{sec:results-perceptual-results}

  \begin{JA}
    対応あり順序尺度データに対しWilcoxon符号順位検定（両側）を使用し，6項目に対しHolm法による多重比較補正を適用した．効果量は$r = Z / \sqrt{N}$で算出した．Fig.~\ref{fig:user-study}(a)に結果を示す．6項目中4項目（Q1, Q3, Q5, Q6）でDELUGEが有意に優位であり（$r = .62$--.80），特にQ3（低遅延）とQ6（全体満足）は$r = .80$と大きな効果量を示した．Q2（視覚的乱れ）とQ4（リアルさ）には有意差がなく，知覚品質差の主因がレイテンシーであることを示唆する．主観的選好では16名全員がDELUGEを選択した（二項検定$p < .001$）．
  \end{JA}
  \begin{EN}
    We used the Wilcoxon signed-rank test (two-tailed) for paired ordinal data with Holm correction across six items; effect size is reported as $r = Z / \sqrt{N}$. Fig.~\ref{fig:user-study}(a) shows the results. Four of the six items (Q1 Smoothness, Q3 Low Latency, Q5 Consistency, Q6 Overall Satisfaction) showed a significant advantage for DELUGE ($r = .62$--.80), with Q3 and Q6 reaching $r = .80$. Q2 (Visual Distortion) and Q4 (Realism) showed no significant difference, suggesting that latency rather than rendering quality is the primary driver of the observed perceptual differences. All 16 participants preferred DELUGE in subjective preference (binomial test, $p < .001$).
  \end{EN}

  \begin{JA}
    Mann-Whitney U検定による順序効果の分析では全6項目で有意差がなく（$p > .13$），カウンターバランスが適切に機能していた．Kruskal-Wallis検定による会場効果でも全6項目で有意差がなく（$p > .11$），結果の会場間ロバスト性が確認された．
  \end{JA}
  \begin{EN}
    Mann-Whitney U tests for order effects showed no significant differences for any of the six items ($p > .13$), confirming effective counterbalancing. Kruskal-Wallis tests for venue effects likewise revealed no significant differences ($p > .11$), supporting the robustness of the results across venues.
  \end{EN}

  \begin{figure}[t]
    \centering
    \includegraphics[width=1\linewidth]{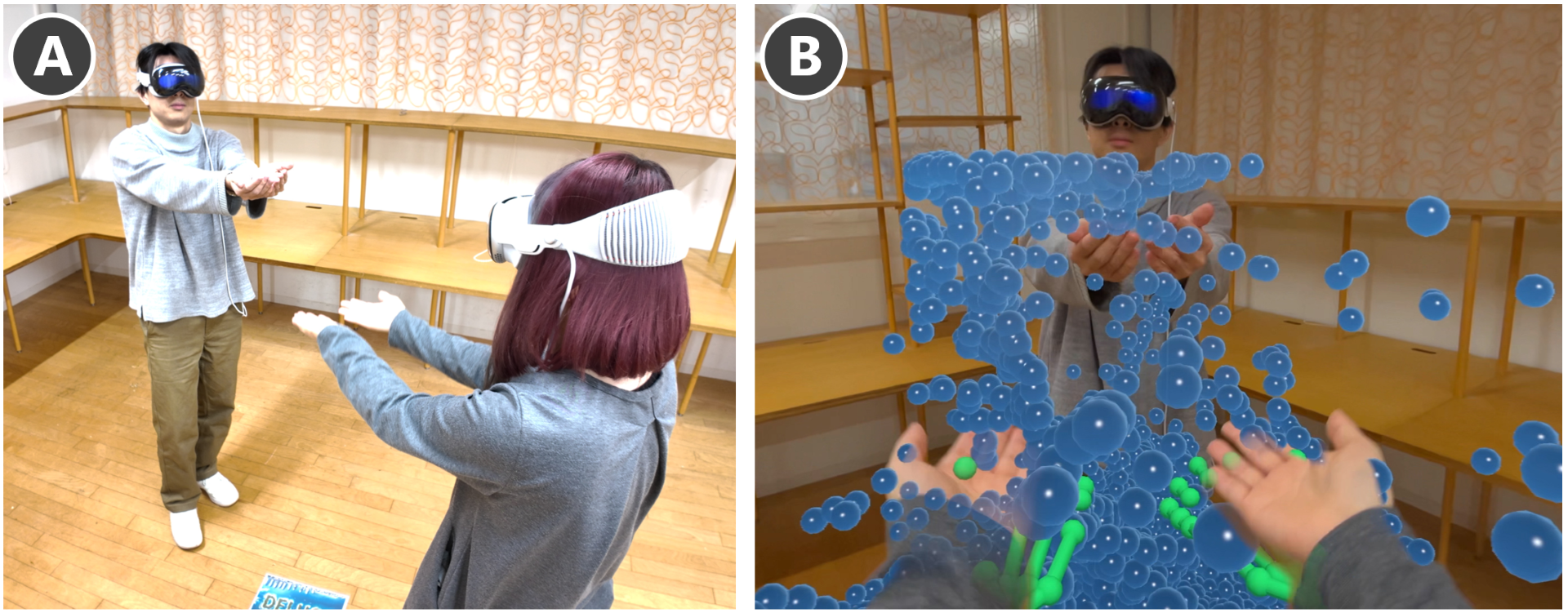}
    \caption{Collaborative task experiment on Apple Vision Pro. (A)~Two participants wearing Vision Pro headsets face each other and simultaneously interact with the same streaming fluid simulation via hand tracking. (B)~First-person view from one participant's headset: both users' hands (green skeleton overlays) manipulate the shared fluid particles in real time, with each user's actions immediately visible to the other.}
    \label{fig:user-photo}
  \end{figure}

  \subsection{Collaborative Task Evaluation}\label{sec:results-collab}

  \subsubsection{Experimental Design}

  \begin{JA}
    同一の16名が，WebSocket経由で接続された2台のクライアント上で実験者とペアとなり，同一の流体に同時に介入する協調タスクを実施した．独立変数はストリーミングシステム条件（DELUGE vs.\ Control，知覚評価と同一構成）であり，従属変数としてQ1: リアルタイム反映，Q2: 視覚的影響，Q3: 遅延知覚（逆転項目），Q4: 状態の一貫性，Q5: やり取り感覚，Q6: 操作への影響，Q7: 意図理解，Q8: 共遊感の8項目を7段階リッカート尺度で評価した（詳細な質問文は補足資料Sec.~\ref{app:questionnaire-details}参照）．
  \end{JA}
  \begin{EN}
    The same 16 participants performed a collaborative task in which the participant and an experimenter simultaneously interacted with the same fluid simulation on two clients connected via WebSocket (Fig.~\ref{fig:user-photo}). The independent variable was the streaming system condition (DELUGE vs.\ Control, identical configurations to the perceptual evaluation). Participants rated eight items on a 7-point Likert scale: Q1~Real-time Reflection, Q2~Visual Influence, Q3~Perceived Delay (reverse-scored), Q4~State Consistency, Q5~Interaction Sense, Q6~Action Influence, Q7~Intent Recognition, and Q8~Co-play Sense (see Sec.~\ref{app:questionnaire-details} for the full item wording).
  \end{EN}

  \subsubsection{Results}\label{sec:results-collab-results}

  \begin{JA}
    Fig.~\ref{fig:user-study}(b)に結果を示す．時間的応答性に関わるQ1--Q3はすべて大きな効果量で有意であった（$r = .81$--.83）．一方，高次社会的知覚（Q4--Q8: 状態一貫性，やり取り感覚，操作影響，意図理解，共遊感）には有意差がなかった．Q5（やり取り感覚）はHolm補正前$p = .051$で境界的であるが，順序尺度データに対するWilcoxon検定の結果を採用し保守的に非有意と判定した．主観的選好ではDELUGE 12名，差なし4名，Control 0名であった（二項検定$p < .001$）．
  \end{JA}
  \begin{EN}
    Fig.~\ref{fig:user-study}(b) shows the results. Q1--Q3, relating to temporal responsiveness, were all significant with large effect sizes ($r = .81$--.83). In contrast, higher-order social items Q4--Q8 (state consistency, interaction sense, action influence, intent recognition, co-play) showed no significant differences. Q5 (Interaction Sense) was borderline ($p = .051$ before Holm correction); we report the Wilcoxon result as the conservative choice for ordinal data. Subjective preference was DELUGE: 12, no difference: 4, Control: 0 (binomial test, $p < .001$).
  \end{EN}

  \begin{JA}
    Q3（遅延知覚）にのみ有意な順序効果が観察された（Mann-Whitney U, $p = .010$）．DELUGE先行群の差分がより大きい（対比効果）が，両群とも群内では有意であり，効果の方向は順序に依存しない．会場効果の分析ではQ2, Q5, Q7に有意な会場差が見られた（Kruskal-Wallis, $p < .04$）：一部の会場ではControl条件でも協調スコアが高く維持されたのに対し，他の会場ではDELUGE優位の差が拡大した．
  \end{JA}
  \begin{EN}
    A significant order effect was observed only for Q3 (Perceived Delay; Mann-Whitney U, $p = .010$): the DELUGE-first group perceived a larger contrast, but both groups showed significant within-group differences, so the direction of the effect is order-independent. Venue analysis revealed significant effects for Q2, Q5, and Q7 (Kruskal-Wallis, $p < .04$): one venue maintained high collaborative scores even under the Control condition, while the other venues showed larger DELUGE advantages.
  \end{EN}

  \section{Discussion}\label{sec:discussion}

  \subsection{Rate--Distortion Characteristics and Design Trade-offs}\label{sec:discussion-rd}

  \begin{JA}
    DELUGEは汎用点群圧縮（G-PCC，Draco）とは異なり，RD効率よりもデコードレイテンシを優先する設計である．Fig.~\ref{fig:ablation-rd}およびTable~\ref{tab:rd-comparison}が示すように，DELUGEは評価ビットレート全域（1.8--9.2\,bppf）でG-PCCと同等以上のPSNRを達成し，高ビットレート域ではG-PCCが評価レンジ内で${\sim}95$\,dB程度に留まるのに対しDELUGEは${\sim}100$\,dB（実質ロスレス）まで到達する．    この差はI/P-frame構造と粒子IDベースのデルタ符号化によるもので，P-frame bppfがI-frame bppfの15--30\%に圧縮される．一方，Dracoはintra単体のRD効率では最も優れるが，符号化時間が65Kを超えると60\,fps予算を超過する（Table~\ref{tab:processing_time}）．DELUGEは復号時間をG-PCCの約$1/20$に抑えつつ（Table~\ref{tab:processing_time}），実用的なRD効率を維持している．

    本手法の利得は3段階に分離できる．(i)~ME-free・ID保存レジームというアーキテクチャ的洞察：動き推定を構造的に排除することが実時間符号化を可能にする前提であり，これを許容するデータ領域の特定自体が貢献である．(ii)~(i)を固定した下での速度適応配分：同一ベース量子化（$B{=}12$）でD1 PSNR $+0.18$\,dB・キーフレームbppf $-19.0\%$（Table~\ref{tab:ablation}；全フレームRDカーブでのBD-PSNRは$+0.5$\,dB）．(iii)~(i)(ii)を固定した下での軸分離ANS：同一PSNRでbppf $-16.0\%$．(ii)(iii)は(i)を固定して計測されているため，ID前提そのものには吸収されない．IDが寄与しないI-only符号化ではG-PCC/Dracoに対し$+29.5\%/+27.0\%$のBD-Rate劣後（Table~\ref{tab:rd-comparison}）となるが，これは静的圧縮効率を犠牲に時間的コヒーレンスを活用する意図的な設計トレードオフを裏付けるものである．  \end{JA}
  \begin{EN}
    Unlike general-purpose point cloud codecs such as G-PCC and Draco, DELUGE prioritizes decoding latency over RD efficiency. As Fig.~\ref{fig:ablation-rd} and Table~\ref{tab:rd-comparison} show, DELUGE matches or exceeds G-PCC in PSNR across its entire evaluated bitrate range (1.8--9.2\,bppf); at higher rates, G-PCC stays around ${\sim}95$\,dB within the evaluated range whereas DELUGE reaches ${\sim}100$\,dB (near-lossless), owing to the I/P-frame structure and particle-ID-based delta coding that compresses P-frame bppf to 15--30\% of I-frame bppf.     Draco achieves the best intra-frame RD efficiency among the three codecs but its encoding exceeds the 60\,fps budget beyond 65K (Table~\ref{tab:processing_time}). DELUGE reduces decoding time to roughly $1/20$ of G-PCC's (Table~\ref{tab:processing_time}) while maintaining practical RD efficiency.

    The gains decompose into three separable contributions. (i)~The architectural insight of the ME-free, ID-preserving regime: structurally eliminating motion estimation is what makes real-time coding possible in the first place, and identifying the data domain that admits this regime is itself a contribution. (ii)~With (i) fixed, velocity-adaptive allocation adds $+0.18$\,dB D1 PSNR at the same base quantization ($B{=}12$) and cuts keyframe bppf by $19.0\%$ (Table~\ref{tab:ablation}; $+0.5$\,dB BD-PSNR on the full-frame RD curve). (iii)~With (i) and (ii) fixed, axis-separated ANS cuts bppf by $16.0\%$ at equal PSNR. Because (ii) and (iii) are measured under a fixed (i), they are not absorbed by the ID premise. For intra-only coding, where IDs contribute nothing, DELUGE trails G-PCC/Draco by $+29.5\%/+27.0\%$ BD-Rate (Table~\ref{tab:rd-comparison}), corroborating the intentional design trade-off of sacrificing static coding efficiency to exploit temporal coherence.  \end{EN}

  \subsection{Decoding Speed and Real-time Feasibility}\label{sec:discussion-realtime}

  \begin{JA}
    Table~\ref{tab:processing_time}から，60\,fps制約（16.67\,ms/frame）に対する各コーデックの実現可能性を議論する．G-PCCは全粒子数で符号化・復号ともに制約を大幅に超過し，CABACの逐次処理がボトルネックとなる．Dracoは復号は全粒子数で制約内（3.4--13.6\,ms）だが，符号化は131K以上で予算を超過する（20.0--41.0\,ms）．    DELUGEは，全フレームの98.3\%を占めるP-frameにおいて262Kでも符号化6.6\,ms・復号5.5\,msと制約内に収まる．I-frame（60フレームに1回）は復号8.1\,msで制約内である一方，符号化は16.2\,msとフレーム周期とほぼ同等であり単発ではわずかに超過しうる；ただしこれは60フレームに1回に限られ，フレーム平均の符号化時間は6.7\,msと予算内であり，サーバのドリフト補正タイマー（Sec.~\ref{sec:architecture}）が単発の超過を吸収する．Apple Vision Pro実機でもDELUGEの復号はI-frame 2.5\,ms / P-frame 1.6\,msであり，90\,fps制約（11.1\,ms）を大きく下回る（Table~\ref{tab:processing_time}）．なお，符号化・伝送・復号を含む配備システム全体のエンドツーエンド遅延は，ユーザ実験セッション全体の中央値でDELUGE 71.5\,ms，Control条件（TMC13ベース）1065.1\,msであった（Sec.~\ref{sec:discussion-subjective}）．    スケーラビリティの観点では，符号化時間は粒子数にほぼ線形にスケールし（65K→262Kの4倍に対し平均1.7→6.7\,ms，約3.9倍），より広い空間カバレッジは$D_{\max}$の増加によって吸収できる．ただし量子化値の16-bit表現（$B \le 16$）により，カバレッジと要求セルサイズの比が$2^{16}$を超える構成では精度が制約される．  \end{JA}
  \begin{EN}
    Table~\ref{tab:processing_time} shows that G-PCC exceeds the 60\,fps budget (16.67\,ms) for both encoding and decoding at all particle counts, with CABAC's sequential processing as the bottleneck. Draco decodes within budget at all particle counts (3.4--13.6\,ms), but its encoding exceeds the budget beyond 65K (20.0--41.0\,ms).     DELUGE stays within the budget for P-frames---98.3\% of all frames at keyframe interval 60---with 6.6\,ms encoding and 5.5\,ms decoding even at 262K. For I-frames (once every 60 frames), decoding takes 8.1\,ms and remains within budget, while encoding takes 16.2\,ms, on par with the frame period and occasionally exceeding it; this is limited to one frame in 60, the per-frame average encoding time stays at 6.7\,ms, and the server's drift-correcting timer (Sec.~\ref{sec:architecture}) absorbs the isolated overrun. On the Apple Vision Pro device, DELUGE decoding takes 2.5\,ms for I-frames and 1.6\,ms for P-frames, well within the 90\,fps budget of 11.1\,ms (Table~\ref{tab:processing_time}). End-to-end latency of the deployed system---including encoding, transport, and decoding---had a pooled median of 71.5\,ms for DELUGE versus 1065.1\,ms for the TMC13-based Control condition across the user-study sessions (Sec.~\ref{sec:discussion-subjective}).     
    In terms of scalability, encoding time scales nearly linearly with particle count (mean 1.7$\to$6.7\,ms for the $4\times$ increase from 65K to 262K, ${\sim}3.9\times$), and larger spatial coverage is absorbed by increasing $D_{\max}$. The 16-bit representation of quantized values ($B \le 16$), however, bounds the achievable precision once the ratio of coverage to the required cell size exceeds $2^{16}$.  \end{EN}

  \subsection{Influence of Dataset Characteristics}\label{sec:discussion-datasets}

  \begin{JA}
    MLS-MPM流体シミュレーションデータは全フレームで粒子数が一定であるため，null点処理が不要であり，DELUGEのI/P-frame構造が最も効果的に機能する条件となる．粒子数の異なる2条件（65Kおよび262K）にわたる評価により，スケーラビリティの分析が可能となる．I/P-frame構造は時間的コヒーレンスの強いシーケンスほど有効に機能するため，粒子密度および運動量の分布がP-frameの圧縮効率に影響を与える．

    RDスイープの結果から，P-frame bppfはI-frame bppfの15--30\%であり，10シーケンス間のP-frame bppfのばらつきは極めて小さい（標準偏差$< 0.5\%$）．これはMLS-MPMシミュレーションにおいて，異なるランダムシードによる初期配置の違いにかかわらず，フレーム間の粒子移動量の統計的性質が安定していることを示す．
    スケーラビリティも良好で，粒子数が4倍（65K→262K）に増加してもbppf増は${\sim}2$倍にとどまる（octreeヘッダの固定コストが按分されるため，Fig.~\ref{fig:ablation-rd}）．
  \end{JA}
  \begin{EN}
    The MLS-MPM fluid simulation data maintains a constant particle count across all frames, eliminating the need for null-point handling and providing ideal conditions for DELUGE's I/P-frame structure. Evaluation across two particle counts (65K and 262K) enables scalability analysis. The I/P-frame structure becomes more effective as temporal coherence increases; particle density and motion magnitude distributions directly influence P-frame compression efficiency.

    The RD sweep results show that P-frame bppf is 15--30\% of I-frame bppf, with negligible variance across the 10 sequences (standard deviation $< 0.5\%$). This indicates that the statistical properties of inter-frame particle displacements remain stable regardless of the random seed used for initial particle placement in the MLS-MPM simulation.
    Scalability is also favorable: a 4$\times$ particle count increase (65K to 262K) yields only ${\sim}2\times$ bppf increase at matched PSNR, as the fixed octree header is amortized over more particles (Fig.~\ref{fig:ablation-rd}).
  \end{EN}

  \subsection{Relationship Between Subjective Scores and Latency}\label{sec:discussion-subjective}

  \begin{JA}
    サーバーログから各セッションのレイテンシーメトリクス（encode/network/decode/E2E）を抽出し，主観スコアとの相関を分析した．実験セッション全体でのE2E中央値はDELUGEが71.5\,ms，Control条件が1065.1\,ms（約15倍差）であった．Control条件のボトルネックはエンコード遅延であり，\texttt{tmc3}リファレンスソフトウェアをsubprocess経由で呼び出す配備形態に由来するラッパーのファイルI/O・シリアライズおよびプロセス起動のオーバーヘッドをアルゴリズムコストに加えて含む（段階別内訳はSec.~\ref{sec:results-perceptual}，コーデック単体のFFI計測はTable~\ref{tab:processing_time}参照）．したがって本節の比較は2つの配備されたシステム条件間の比較である．ただし，以下の主観スコアとの相関分析は「レイテンシー差が体験に与える影響」を検証するものであり，レイテンシー差の原因に依存しない．  \end{JA}
  \begin{EN}
    We extracted per-session latency metrics (encode, network, decode, and end-to-end, \ie, E2E) from server logs and analyzed their correlation with subjective scores. The pooled median E2E latency across all study sessions was 71.5\,ms for DELUGE and 1065.1\,ms for the Control condition (a ${\sim}15\times$ difference). The bottleneck in the Control condition was encoding latency: the \texttt{tmc3} reference software was invoked as a blocking subprocess, so the measured latency includes the wrapper's file I/O, serialization, and process startup overhead in addition to the algorithmic cost (see Sec.~\ref{sec:results-perceptual} for the per-stage breakdown; the codec-only comparison is provided by the FFI measurements in Table~\ref{tab:processing_time}). The comparison in this section is thus between two deployed system conditions. The correlation analysis below examines how latency differences affect the user experience, and its validity does not depend on the source of the latency gap.  \end{EN}

  \begin{JA}
    全コーデック横断のSpearman相関ではE2E中央値と遅延関連項目（知覚Q1, Q3, Q6，協調Q1--Q3）に有意な相関（$\rho = -0.43$〜$+0.58$, $p < .05$）が見られたが，レンダリング品質項目（知覚Q2, Q4）や高次協調項目（Q4--Q8）では相関がなかった．しかし条件内（DELUGE内またはControl内）ではこれらの相関がほぼ消失した．これは全体相関が条件間のレイテンシー差（71.5\,ms vs 1065.1\,ms）に起因することを示し，知覚的レイテンシー閾値の存在を示唆する：DELUGEの範囲（4--483\,ms）とControl条件の範囲（491--2104\,ms）は閾値の両側に位置し，条件内のばらつきは閾値をまたがない．
  \end{JA}
  \begin{EN}
    Across both codecs, Spearman correlations between median E2E latency and subjective scores were significant for latency-sensitive items (perceptual Q1, Q3, Q6 and collaborative Q1--Q3; $\rho = -0.43$ to $+0.58$, $p < .05$) but not for rendering quality items (perceptual Q2, Q4) or higher-order collaborative items (Q4--Q8). Critically, within each condition these correlations vanished almost entirely, indicating that the overall correlations were driven by the between-condition latency gap (71.5\,ms vs.\ 1065.1\,ms) rather than within-condition variability. This suggests the existence of a perceptual latency threshold: DELUGE's range (4--483\,ms) and the Control condition's range (491--2104\,ms) lie on opposite sides of this threshold, while within-condition variance does not cross it.
  \end{EN}

  \begin{JA}
    会場間差異の分析では，DELUGE条件でV3会場がV1/V2の約2.4倍のレイテンシーを示し（Kruskal-Wallis $p = .002$），ネットワーク環境差の存在が示唆された．Control条件の協調スコアはV1会場で高く維持された一方，V2/V3会場では低下しており，参加者間の社会的親密度の差がレイテンシーの知覚的影響を調節している可能性がある．
  \end{JA}
  \begin{EN}
    Venue analysis revealed that V3 showed 2.4$\times$ higher DELUGE latency than V1/V2 (Kruskal-Wallis, $p = .002$), attributable to network infrastructure differences. Control-condition collaborative scores remained high at Venue V1 but dropped at Venues V2/V3, possibly reflecting a moderating effect of pre-existing social familiarity between participants on the perceived impact of latency.
  \end{EN}

  \section{Limitations and Future Work}\label{sec:limitations}

  \subsection{Codec Limitations}

  \begin{JA}
    本研究の貢献は，シミュレータが粒子IDを保持する——すなわちフレーム間で安定な点対応が存在する——インタラクティブ・シミュレーションストリームにスコープされる．第一に，DELUGEはフレーム間で粒子数が一定でIDが保存されることを前提としており，可変点数やトポロジ変化（分裂・合流）を伴うデータ，およびIDを持たない実計測データへの対応には，キーフレーム再初期化・フレーム間対応探索・null-point処理が必要であり，今後の課題である．実計測点群での予備評価（Sec.~\ref{sec:results-captured}）が示すとおり，このボトルネックは圧縮アルゴリズムではなくID割当てにあり，本手法のGlobal ID（Morton順序）を対応探索の高速化に再利用することが将来研究の手掛かりとなる．  \end{JA}
  \begin{EN}
    The contribution of this work is scoped to interactive simulation streams in which the simulator preserves particle IDs---\ie, streams with stable frame-to-frame point correspondence. First, DELUGE assumes a fixed particle count with preserved IDs across frames; supporting variable point counts, topology changes (splitting, merging), and captured data without IDs would require keyframe re-initialization, inter-frame correspondence search, and null-point handling, which remain future work. As the preliminary evaluation on captured point clouds shows (Sec.~\ref{sec:results-captured}), this bottleneck is ID assignment---not the compression algorithm; reusing our Global ID (Morton ordering) to accelerate correspondence search is a natural starting point.  \end{EN}

  \subsection{User Study Limitations}

  \begin{JA}
    ユーザ実験はLAN環境で実施しており，WAN環境でのパケットロスやジッターの影響は評価していない．協調タスクのペアは常に実験者＋被験者であり，一部の会場では参加者と実験者が事前に面識があったため，社会的親密度の差が協調評価の会場差に寄与している可能性がある．
  \end{JA}
  \begin{EN}
    User experiments were conducted over a LAN; the effects of packet loss and jitter in WAN settings remain unevaluated. Collaborative pairs always consisted of one experimenter and one participant; at one venue, participants had pre-existing familiarity with the experimenter, and this social proximity may have contributed to the observed venue effects on collaborative scores.
  \end{EN}

  \subsection{Future Directions}

  \begin{JA}
    フレーム間予測の高度化（動きベクトルによる予測残差の削減），適応的キーフレーム間隔，およびマルチレゾリューション伝送による帯域適応は，圧縮効率のさらなる改善に向けた重要な方向性である．また，アバター姿勢や環境オブジェクトなど，粒子シミュレーション以外の3次元点群への応用も検討に値する．
  \end{JA}
  \begin{EN}
    Improving inter-frame prediction through motion vectors, adapting the keyframe interval dynamically, and introducing multi-resolution transmission for bandwidth adaptation are important directions for further compression gains. Extending DELUGE to other 3D point cloud domains such as avatar poses and environmental objects is also worth investigating.
  \end{EN}

  \section{Conclusion}\label{sec:conclusion}

  \begin{JA}
    本研究では，マルチユーザVR環境における大規模粒子シミュレーションのリアルタイム双方向インタラクションを実現するストリーミング圧縮アーキテクチャDELUGEを提案した．
    DELUGEの核心は，物理シミュレーション粒子に固有の特性を圧縮の各段階で体系的に活用する3つの着想にある．第一に，高速粒子の知覚的マスキング効果を利用した速度適応型ビット配分により，知覚品質を維持しつつキーフレームのビット予算とレイテンシを削減した．
    第二に，軸分離型エントロピー符号化で，軸方向の分布偏りの個別捕捉と3軸並列復号を同時に実現した．第三に，平坦逆量子化LUTにより木構造の探索的アクセスを排して$O(N)$アクセスを可能とし，SIMD自動ベクトル化が有効な単一ループでの高速復号を実現した．
  \end{JA}
  \begin{EN}
    We have presented DELUGE, a streaming compression architecture that enables real-time bidirectional interaction with large-scale particle simulations in multi-user VR environments.
    The core of DELUGE lies in three ideas that systematically exploit properties specific to physics-simulation particles at each stage of compression. First, velocity-adaptive bit allocation takes advantage of the perceptual masking of fast-moving particles to reduce the keyframe bit budget and latency while preserving perceived quality    Second, axis-separated entropy coding captures per-axis distribution biases independently and enables three-way parallel decoding. Third, the flat inverse-quantization LUT replaces tree traversal with $O(N)$ array access, enabling single-loop decoding amenable to SIMD auto-vectorization.
  \end{EN}

  \begin{JA}
    動的点群および流体シミュレーションデータセットを用いた評価において，DELUGEはG-PCC (TMC13)に対し約$20\times$高速な復号，Dracoに対し$6\times$高速な符号化を達成した（全コーデックFFI in-process計測）．

    Dracoに対するBD-Rateコストはデコードレイテンシを優先する意図的な設計判断であり，WebAssemblyおよびネイティブプラットフォームの両方で60\,fpsリアルタイムストリーミングの要件を満たす．Apple Vision Pro上での被験者内比較実験（$N = 16$）では，DELUGEベースのストリーミングシステムがTMC13ベースの配備システムに対し，配備されたインタラクション条件間の比較として遅延関連の知覚・協調項目で有意に優れ（Wilcoxon, $r = .62$--$.83$），E2Eレイテンシーの低減がリアルタイム協調体験の実現に直結することを実証した．  \end{JA}
  \begin{EN}
    Evaluation on dynamic point cloud and fluid simulation datasets showed that DELUGE achieves approximately $20\times$ faster decoding than G-PCC (TMC13) and $6\times$ faster encoding than Draco when measured in-process via FFI.

    The BD-Rate cost relative to Draco is an intentional design trade-off that prioritizes decoding latency, and DELUGE satisfies the 60\,fps real-time streaming requirement on both WebAssembly and native platforms. A within-participants experiment on Apple Vision Pro ($N = 16$), comparing the two deployed interaction conditions, demonstrated that the DELUGE-based streaming system significantly outperforms the TMC13-based deployed system on latency-related perceptual and collaborative items (Wilcoxon signed-rank, $r = .62$--.83), confirming that reducing E2E latency directly enables real-time collaborative experiences.  \end{EN}

  \acknowledgments{This work was partially supported by JST Moonshot Research \& Development Program Grant Number JPMJMS2013 and JST ASPIRE Grant Number JPMJAP2327.}

  \bibliographystyle{abbrv-doi-hyperref}
  \bibliography{template,supplementary}

\clearpage

\appendix
\crefalias{section}{appendix}
\setcounter{figure}{0}
\setcounter{table}{4}
\setcounter{equation}{11}
\renewcommand{\thefigure}{S\arabic{figure}}
\twocolumn[\begin{center}
\vspace{1em}
{\sffamily\LARGE\textbf{Supplementary Material for ``DELUGE: \\
Decomposed Entropy-coded Live Unstructured Geometry Exchange \\
for Real-time Particle Streaming''}}
\vspace{1em}
\end{center}
]

\section{Additional Rate--Distortion Result}\label{app:rd-65k}

\begin{figure}[t]
\centering
\includegraphics[width=\linewidth]{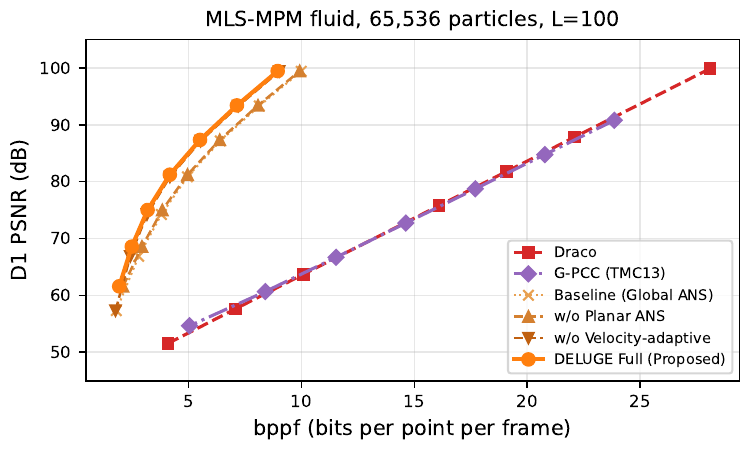}
\caption{Rate--distortion curves on MLS-MPM fluid simulation at 65K particles ($L{=}100$; 10-sequence average over all 600 frames; D1 PSNR every 10th frame; $D_{\max}{=}7$, keyframe interval 60).}
\label{fig:rd-65k}
\end{figure}

Figure~\ref{fig:rd-65k} complements the 262K-particle result in main-text Fig.~\ref{fig:ablation-rd} with the corresponding 65K-particle curves. Both particle counts show the same codec ordering; the improved RD efficiency at 262K results from amortizing the fixed octree-header cost over more particles.

\section{Codec Parameters}\label{app:codec-params}

Table~\ref{tab:codec-params} lists the tunable parameters of DELUGE for reproducibility.

\begin{table}[h]
\centering
\caption{DELUGE tunable parameters.}
\label{tab:codec-params}
\scriptsize
\setlength{\tabcolsep}{4pt}
\renewcommand{\arraystretch}{1.0}
\begin{tabular}{@{}l l l l@{}}
\toprule
\textbf{Parameter} & \textbf{Symbol} & \textbf{Value} & \textbf{Explored range} \\
\midrule
Keyframe interval        & ---            & 60 & 30--120 \\
Max octree depth         & $D_{\max}$     & 7 & 4--8 \\
Base quantization bits    & $B$            & 12 & 8--14 (RD sweep) \\
Velocity thresholds      & $\alpha_1, \alpha_2$ & 0.05, 0.10 & 0--0.5, 0--1.0 \\
History window           & $w$            & 5 & 3--10 \\
Frequency table size     & $M$            & 4096 ($=2^{12}$) & $2^{10}$--$2^{13}$ \\
ANS state lower bound    & $L$            & 32768 ($=2^{15}$) & $2^{14}$--$2^{16}$ \\
\bottomrule
\end{tabular}\end{table}

\section{Octree Construction Details}\label{app:octree-details}

\subsection{Bounding Box Construction}\label{app:bounding-box}

Given an input particle set $\mathcal{P} = \{\mathbf{p}_i\}_{i=1}^{N}$ ($\mathbf{p}_i \in \mathbb{R}^3$), we construct a cubic bounding box that serves as the root node of the octree. The per-axis extrema, center $\mathbf{c}$, and half-extent $E$ are computed as:
\begin{align}
  \mathbf{p}_{\min} &= \bigl(\min_{i} p_{i,x},\;\min_{i} p_{i,y},\;\min_{i} p_{i,z}\bigr) \\
  \mathbf{p}_{\max} &= \bigl(\max_{i} p_{i,x},\;\max_{i} p_{i,y},\;\max_{i} p_{i,z}\bigr) \\
  \mathbf{c} &= \tfrac{\mathbf{p}_{\min} + \mathbf{p}_{\max}}{2} \\
  E &= \max\!\bigl(\tfrac{p_{\max,x}-p_{\min,x}}{2},\;
       \tfrac{p_{\max,y}-p_{\min,y}}{2},\;
       \tfrac{p_{\max,z}-p_{\min,z}}{2},\; \epsilon\bigr)
\end{align}
The maximum of the three half-widths is used because the octree subdivides space isotropically, requiring a cubic (not rectangular) domain. $\epsilon = 10^{-3}$ prevents degenerate zero-extent boxes when all particles coincide.

\subsection{Occupancy Mask}\label{app:occupancy-mask}

The occupancy mask is logically a bit array of $T = (8^{D_{\max}+1} - 1) / 7$ bits, one bit per node in the full octree. Bits corresponding to the Global IDs of active nodes (leaves containing at least one particle) are set to 1; all others are 0. This mask is transmitted only in keyframes. On the wire, the sparse bit array is serialized as a delta-varint compressed list of the active Global IDs in ascending order, which is compact because active IDs are sparse ($N_{\mathrm{leaf}} \ll T$) and sorted. Since the Global ID (Eq.~(2)) increases with depth level and follows Morton order within each level, ascending Global ID order coincides with a breadth-first (BFS) traversal of the octree. On the decoder side, parsing the delta-varint list---logically equivalent to scanning the bit array in BFS order---recovers the active nodes, and the bounds of each node are geometrically reconstructed from its Global ID using the inverse of Eq.~(2) in the main text.
\subsection{Morton Code Equivalence}\label{app:morton-equivalence}

The octree path index $k = \sum_{d'=0}^{d_i-1} b_{i,d'} \cdot 8^{d_i - 1 - d'}$ is mathematically equivalent to a Morton code (Z-order curve)~\cite{morton1966}. Each level's octant number $b_{i,d} = b_x + 2b_y + 4b_z$ interleaves one bit from each axis; concatenating these from root to leaf produces the bit-interleaved Morton code of the particle's normalized coordinates. Morton-code-based spatial ordering is a foundational technique used in linear quadtrees~\cite{gargantini1982}, spatial data structure taxonomies~\cite{samet2006}, GPU BVH construction~\cite{karras2012}, and octree geometry coding in MPEG G-PCC.

\subsection{Octree vs.\ KD-Tree}\label{app:kdtree}

Two structural properties favor the octree over a KD-tree for our codec. First, octree cells are isotropic cubes whose bounds are fully determined by the Global ID (Eq.~(2)), so leaf depth is a direct proxy for world-space resolution---the property that depth-dependent quantization (Eq.~(5)) relies on---and no per-leaf bounds need to be transmitted. KD-tree leaves are data-dependent anisotropic rectangles: the depth--resolution correspondence breaks down, and each leaf requires explicit AABB metadata ($6 \times$ \texttt{f32} $=$ 24 bytes per leaf), which inflates keyframe headers as the leaf count grows. Second, the binary split requires roughly $3\times$ the depth of an octree for the same leaf count, and because the per-leaf bit budget follows $b_\ell = \max(1, B - D_L)$ (Eq.~(5)), deep KD-trees collapse the effective coordinate precision.

A preliminary comparison (8{,}192 particles, 10 delta frames, root-bits sweep 6--10) confirms this: against the Octree $d{=}3$ anchor, the best KD-tree configuration ($d{=}3$) incurs $+17.4\%$ overall BD-Rate ($+19.4\%$ at $d{=}4$; $+40.2\%$ on keyframes alone), and at $d{=}12$ (4{,}096 leaves) the per-leaf AABB metadata by itself exceeds the raw data size (135.9 bppf keyframes).

\section{Entropy Coding Details}\label{app:entropy-details}

\subsection{Semi-Static rANS Encoding}\label{app:semi-static-ans}

We use rANS in a semi-static mode~\cite{moffat1995semi_static}. The entire frame is scanned to collect symbol occurrence frequencies, and the resulting probability table is used for batch encoding. This probability table is then transmitted alongside the compressed data.

The semi-static approach is robust against abrupt changes in particle motion. Adaptive methods, which update the probability model sequentially as CABAC does, can suffer a temporary but significant drop in coding efficiency when the simulation state changes rapidly. In contrast, the semi-static approach builds a probability table that reflects the statistics of the entire frame, adapting immediately to abrupt distribution changes and maintaining stable coding efficiency.

Because quantized values with $B = 12$ occupy $[0, 4095]$, each value is first split into its high and low bytes, which are entropy-coded independently (for $b_\ell \le 8$ the high bytes are all zero and compress to negligible size). Each rANS stream therefore operates on a 256-symbol alphabet. Note that the alphabet size (256) and the normalized frequency sum $M$ are distinct quantities: the former is the number of codable symbols, the latter the granularity of the ANS state space. The frequency table is normalized to a power of two, $M = 2^{12} = 4096$, and the ANS state lower bound is set to $L = 2^{15} = 32768$. For each byte stream of the symbol sequence $\{s_j\}_{j=1}^{3N}$ obtained from $N$ particles, encoding proceeds as follows: (1)~the occurrence count $c_s$ of each symbol $s \in [0, 255]$ is tallied;(2)~every symbol is assigned a minimum frequency of 1, and the remainder is distributed proportionally to the observed ratios, with rounding errors corrected in descending frequency order to ensure $\sum f_s = M$; (3)~the rANS algorithm compresses the data using this table, processing symbols in reverse order and folding the information of symbol $s$ into the state via $x' = \lfloor x / f_s \rfloor \cdot M + (x \bmod f_s) + F_s$ (where $F_s = \sum_{s'<s} f_{s'}$ is the cumulative frequency), with the lower 8 bits output to a byte stream for renormalization when the state exceeds its upper bound; and finally (4)~a 256-bit Frequency Mask (a fixed 32-byte flag indicating non-zero-frequency symbols) together with the frequency values (16 bits each) of flagged symbols are appended to the packet~\cite{cram31}.

\subsection{ZigZag Encoding and Circular Differencing}\label{app:zigzag-main}

Delta frames encode the difference in quantized coordinates from the preceding frame. With $B = 12$, quantized values lie in $[0, 4095]$. We treat these 4096 values as a ring (mod 4096) and use circular differencing to keep the shortest-path difference within $[-2048, 2047]$:
\begin{equation}
  \delta = \bigl((q^{(t)} - q^{(t-1)} + 2048) \bmod 4096\bigr) - 2048
\end{equation}

The signed difference $\delta$ is then converted to a non-negative integer $z$ via ZigZag encoding~\cite{protobuf_encoding}, which bijects signed integers to non-negative integers as $0 \to 0,\; {-1} \to 1,\; 1 \to 2,\; {-2} \to 3,\; \ldots$:
\begin{equation}
  z = (\delta \ll 1) \oplus (\delta \gg 11)
\end{equation}
The resulting $z$ values concentrate near zero, improving entropy coding efficiency. For keyframes, the quantized coordinates $q_{i,a}$ are used directly as symbols.

\subsection{Frequency Mask}\label{app:freq-mask}

rANS decoding requires the same frequency table used by the encoder. Transmitting all 256 symbol frequencies directly at 16 bits each costs $256 \times 16 = 4{,}096$ bits (512 bytes). We reduce this overhead with a Frequency Mask~\cite{cram31}: a fixed 32-byte (256-bit) bitmask indicates which symbols have $f_s > 0$, and only the frequencies of flagged symbols are transmitted at 16 bits each. If $K$ symbols are present, the table overhead is $256 + 16K$ bits ($32 + 2K$ bytes). In delta frames, ZigZag encoding concentrates difference symbols near zero, so $K \ll 256$ is typical and the savings are substantial.

\section{Packet Format and Decoding Process}\label{app:packet-details}

\subsection{Packet Format}

Packets are designed in a compact binary format that uses bandwidth efficiently and enables low-latency decoding. Table~\ref{tab:packet_format_app} shows the detailed bit-packing layout for keyframes and delta frames.

\begin{table}[h]
\centering
\caption{Packet Format for Keyframe and Delta Frame. $\checkmark$ indicates fields included in both the delta frame and the keyframe; unmarked fields are transmitted only in the keyframe.}
\label{tab:packet_format_app}
\small
\setlength{\tabcolsep}{2pt}
\renewcommand{\arraystretch}{0.95}
\begin{tabular}{@{}l l r c p{0.40\columnwidth}@{}}
\toprule
\textbf{Field} & \textbf{Type} & \textbf{B} & \textbf{Delta} & \textbf{Description} \\
\midrule
Packed Type & \texttt{U8} & 1 & $\checkmark$ & 4-bit encoding type + 4-bit packet type \\
Frame ID & \texttt{U16} & 2 & $\checkmark$ & Sequence index (little endian) \\
Timestamp & \texttt{F64} & 8 & $\checkmark$ & Server timestamp (Unix epoch) \\
Center (x,y,z) & \texttt{F32$\times$3} & 12 &  & Bounding box center \\
Extent & \texttt{F32} & 4 &  & Bounding box half-extent \\
Max Depth & \texttt{U8} & 1 &  & Octree maximum depth \\
Total Nodes & \texttt{U32} & 4 &  & Total nodes in full tree \\
Active Nodes & \texttt{U32} & 4 &  & Active (non-empty) nodes \\
Occ.\ Mask & \texttt{B[]} & var. &  & Occupancy bitmask (BFS order) \\
Prefix Sum & \texttt{U32[]} & var. &  & Cumulative particle counts per active node \\Payload Len & \texttt{U32} & 4 & $\checkmark$ & Byte length of compressed payload \\
Freq Mask & \texttt{U8[32]} & 32 & $\checkmark$ & Presence flags for symbols 0--255 \\
Freq Values & \texttt{U16[]} & var. & $\checkmark$ & Frequencies for present symbols \\
ANS Stream & \texttt{B[]} & var. & $\checkmark$ & rANS-coded residual stream \\
\bottomrule
\end{tabular}
\end{table}

Keyframes include spatial structure information. The Occupancy Mask is logically a bit array of $T = (8^{D_{\max}+1} - 1) / 7$ bits, where $D_{\max}$ is the maximum depth of the octree; the bits corresponding to the active-node Global IDs are set to 1, and all the others are set to 0. On the wire it is serialized as a delta-varint compressed list of the active Global IDs in ascending order, which coincides with BFS order (Morton order within each depth level; Sec.~\ref{app:occupancy-mask}).
The decoder parses this list---logically equivalent to scanning the bit array in BFS order---to reconstruct the active nodes and geometrically recover the bounds of each node from its Global ID. The Prefix Sum stores the cumulative particle counts across active nodes as $K + 1$ values, likewise delta-varint coded. Delta frames omit the spatial structure entirely, relying on the assumption that the structure established by the most recent keyframe has not changed.

\section{Captured-Data Evaluation Details}\label{app:captured-details}

Table~\ref{tab:captured-eval} details the preliminary evaluation on captured point clouds summarized in Sec.~5.2.3 of the main text.

\textbf{Setup.} We use the first 300 frames of each sequence (Thaidancer comprises 300 frames in total): 8iVSLF Thaidancer\_viewdep (vox12)~\cite{krivokuca_8ivslf_2018} and Owlii basketball\_player / dancer (vox11)~\cite{xu_owlii_2017}, at their native per-frame point counts without subsampling. Because these sequences carry no frame-to-frame point correspondence, DELUGE operates all-intra (every frame coded as a keyframe) with $D_{\max}{=}3$ and $B{=}8$, axis-separated Planar rANS; this shallow-octree operating point matches the voxelized data resolution. G-PCC is TMC13 v23.0-rc2 with \texttt{pqs\,=\,0.25}, likewise intra-only. Both codecs run in-process (Rust/C FFI, one call per frame, no subprocess) on Apple M2 Max, following the measurement protocol of main-text Table~3; decoding times are per-frame means over the 300 frames, with data loading excluded from all timed regions. bppf is aggregated over the same 300 frames from the encoded packet sizes, and D1 PSNR is evaluated on every frame with the MPEG \texttt{pc\_error} tool.

\begin{table}[h]
\centering
\caption{Preliminary evaluation on captured point clouds: 8iVSLF Thaidancer and Owlii basketball\_player / dancer (all-intra; DELUGE $D_{\max}{=}3$, $B{=}8$; G-PCC \texttt{pqs\,=\,0.25}; first 300 frames of each sequence).}
\label{tab:captured-eval}
\scriptsize
\setlength{\tabcolsep}{2pt}
\renewcommand{\arraystretch}{0.95}
\begin{tabular}{@{}l r r r r r r@{}}
\toprule
 & & & \textbf{D1 PSNR} & \multicolumn{2}{c}{\textbf{Decode (ms)}} & \\
\cmidrule(lr){5-6}
\textbf{Sequence} & \textbf{Points} & \textbf{bppf} & \textbf{(dB)} & \textbf{DELUGE} & \textbf{G-PCC} & \textbf{Ratio} \\
\midrule
Thaidancer & 3.08M & 15.46 & 59.89 & 56.8 & 1{,}131 & 19.9$\times$ \\
basketball\_player & 2.91M & 15.46 & 60.26 & 55.2 & 1{,}166 & 21.1$\times$ \\
dancer & 2.60M & 15.46 & 60.21 & 47.4 & 1{,}019 & 21.5$\times$ \\
\midrule
Mean & & 15.46 & 60.12 & & & 20.9$\times$ \\
\bottomrule
\end{tabular}
\end{table}

\section{visionOS Client Implementation Details}\label{app:visionos-details}

\subsection{Hand Tracking and Fluid Interaction}

The visionOS client uses ARKit's \texttt{HandTrackingProvider} to obtain the skeleton of both hands at each frame. Each hand consists of 27 joints (wrist, metacarpal, knuckle, intermediate, and tip for each of five fingers, plus two forearm joints). The world-space position of each joint is computed as \texttt{originFromAnchorTransform} $\times$ \texttt{anchorFromJointTransform}.

The joint positions are transmitted to the server as a JSON message at each rendering frame (90\,Hz). Each joint is represented as a four-element tuple $(x, y, z, r)$ encoding its position and collision radius ($r = 8$\,mm). Bones connecting adjacent joints are represented as start--end pairs $(x_a, y_a, z_a, r_a, x_b, y_b, z_b, r_b)$.

On the server, the received skeleton data is transformed into the simulation coordinate system and converted into colliders. Joints become sphere colliders and bones become capsule colliders; these are uploaded to the MLS-MPM GPU kernel every frame. Within the kernel, the signed distance from each particle to the nearest collider is evaluated, and a repulsive force is applied when a particle penetrates a collider surface. This enables real-time fluid interaction in which users can push, scoop, and redirect the fluid with their bare hands.

\subsection{Coordinate Alignment}

In the multi-user setting, the coordinate systems of multiple Vision Pro devices must be aligned with the server's simulation space. We use image-marker-based coordinate alignment via ARKit's \texttt{ImageTrackingProvider}. Each device detects a pre-registered reference image of known physical size and obtains its world-space transform. This transform serves as an anchor from which a model matrix is constructed to map the simulation origin, rotation, and scale into world space. Marker detection is performed once at startup; subsequent tracking relies on the device's spatial tracking system.

On the server, the skeleton-to-collider conversion applies the inverse of the marker transform to convert world coordinates to simulation coordinates. This ensures that skeleton data from different devices is correctly resolved in the same simulation space. In addition to its own skeleton, each client receives the skeleton data of other clients broadcast via the server, and renders remote users' hands. This enables a collaborative experience in which multiple users visually share and simultaneously manipulate the same fluid.

\section{Questionnaire Items}\label{app:questionnaire-details}

\subsection{Perceptual Quality Evaluation (Sec.~\ref{sec:results-perceptual})}
All items were rated on a 7-point Likert scale (1: strongly disagree -- 7: strongly agree).
\begin{itemize}[leftmargin=*]
  \item \textbf{Q1} (Smoothness): The fluid motion appeared smooth and continuous.
  \item \textbf{Q2} (Visual Distortion; reverse-scored): I noticed visual distortion or unnaturalness in the fluid.
  \item \textbf{Q3} (Low Latency): The fluid responded immediately to my hand movements.
  \item \textbf{Q4} (Realism): The fluid behavior looked natural and realistic.
  \item \textbf{Q5} (Consistency): The shape and density of the fluid appeared spatially coherent.
  \item \textbf{Q6} (Overall Satisfaction): Overall, I was satisfied with the visual quality of the fluid.
\end{itemize}

\subsection{Collaborative Task Evaluation (Sec.~\ref{sec:results-collab})}
All items were rated on a 7-point Likert scale (1: strongly disagree -- 7: strongly agree).
\begin{itemize}[leftmargin=*]
  \item \textbf{Q1} (Real-time Reflection): My partner's actions were reflected in the fluid in real time.
  \item \textbf{Q2} (Visual Influence): I could clearly see my partner's influence on the fluid.
  \item \textbf{Q3} (Perceived Delay; reverse-scored): I noticed a noticeable delay in the shared environment.
  \item \textbf{Q4} (State Consistency): The fluid state appeared consistent between both users.
  \item \textbf{Q5} (Interaction Sense): I felt like I was interacting with my partner through the fluid.
  \item \textbf{Q6} (Action Influence): My partner's actions influenced my own actions.
  \item \textbf{Q7} (Intent Recognition): I could understand what my partner was trying to do.
  \item \textbf{Q8} (Co-play Sense): I felt like we were playing together.
\end{itemize}

\end{document}